\documentclass[preprintnumbers,showpacs,floats,twocolumn,prd,aps]{revtex4}
\usepackage{amssymb}
\usepackage[fleqn]{amsmath}
\usepackage{epsfig,hyperref}
\usepackage[svgnames]{xcolor}
\usepackage{pgf,tikz}
\usepackage{epstopdf}
\usepackage[final]{pdfpages}
\DeclareMathOperator{\sech}{sech}

\makeatletter
\DeclareRobustCommand*{\bfseries}{%
  \not@math@alphabet\bfseries\mathbf
  \fontseries\bfdefault\selectfont
  \boldmath
}
\makeatother

\def\be{\begin{equation}}
\def\ee{\end{equation}}
\def\beq{\begin{eqnarray}}
\def\eeq{\end{eqnarray}}

\makeatletter
\renewcommand{\vec}[1]{\mbox{\boldmath$#1$}}
\newcommand{\arXiv}[2][]{\href{http://arxiv.org/abs/#2}{\texttt{arXiv:#2\@ifempty{#1}{}{ [#1]}}}}
\makeatother

\begin{document}

\title{Cosmological dynamics in $f(R)$ gravity}

\author{Jun-Qi Guo}%
\email{jga35@sfu.ca}
\author{Andrei V. Frolov}
\email{frolov@sfu.ca}
\affiliation{
Department of Physics, Simon Fraser University\\
8888 University Drive, Burnaby, BC V5A 1S6, Canada
}%

\date{\today}

\begin{abstract}
In this paper, we study the cosmological viability conditions, the phase-space dynamics, and the cosmological evolution of $f(R)$ gravity.
In contrast to most previous works in the literature, which analyzed the background dynamics of $f(R)$ gravity by means of a dynamical system, we proceed by focusing on the equivalent scalar field description of the theory, which we believe is a more intuitive way of treating the problem.
In order to study how the physical solutions evolve in $f(R)$ cosmology, we explore the cosmological dynamics of a range of $f(R)$ models, including models that yield a large hierarchy of scales and are singularity free. We present generic features of the phase-space dynamics in $f(R)$ cosmology. We study the global structure of the phase space in $f(R)$ gravity by compactifying the infinite phase space into a finite space via the $\text{Poincar\'e}$ transformation.
On the expansion branch of the phase space, the constraint surface has a repeller and a de Sitter attractor; while on the contraction branch, the constraint surface has an attractor and a de Sitter repeller. Generally, the phase currents originate from the repeller and terminate at the corresponding attractor in each space. The trajectories between the repeller and the attractor in the presence of matter density are different from those in the vacuum case.
The phase analysis techniques developed in this paper are very general, and can be applied to other similar dynamical systems.
\end{abstract}
\pacs{04.50.Kd, 05.45.-a, 98.80.-k}% PACS, the Physics and Astronomy
\keywords{}
%\preprint{SCG-2013-03}
\maketitle

%%%%%%%%%%%%%%%%%%%%%%%%%%%%%%%%%%%%%%%%%%%%%%%%%%%%%%%%%%%%%%%%%%%%%%%%%%%%%%%%%%%%%%%%%%%%%%%%%%%%%%%%%%%%%%%%%%%%%%%%%%
%
\section{\label{sec:introduction}Introduction}
%%%%%%%%%%%%%%%%%%%%%%%%%%%%%%%%%%%%%%%%%%%%%%%%%%%%%%%%%%%%%%%%%%%%%%%%%%%%%%%%%%%%%%%%%%%%%%%%%%%%%%%%%%%%%%%%%%%%%%%%%%
The measurements of Type Ia supernovae luminosity distances indicate that the current Universe is undergoing an accelerated
expansion~\cite{Riess:1998cb,  Perlmutter:1998np, Knop, Riess:2004nr, Planck:2013}. The simplest approach to address this issue is to introduce the $\Lambda$CDM model, in which $31.7\%$ of the mass-energy density of the Universe is made up of ordinary matter and dark matter, and the rest is constituted by the cosmological constant, $\Lambda$~\cite{Planck:2013}. The cosmological constant has large negative pressure, and the equation of state ($w \equiv P/\rho$) is equal to $-1$, where $P$ and $\rho$ are the pressure and the energy density of the cosmological constant, respectively. It is the large negative pressure that functions as the repulsive force field against the regular gravity, thus driving the cosmic acceleration. However, the value of the observed cosmological constant is less than the Planck scale by a factor of 120 orders of magnitude~\cite{Carroll}.

Another possibility is that the cosmic speed-up might be caused within general relativity by a mysterious cosmic fluid with negative pressure, which is usually called \lq\lq dark energy\rq\rq.~However, the nature of dark energy is still unknown. Alternatively, the acceleration could be due to purely gravitational effects, i.e. one may consider modifying the current gravitational theory to produce an effective dark energy. A natural approach is to replace the Ricci scalar in the Einstein-Hilbert action with an arbitrary function of the Ricci scalar,
\be S=\frac{1}{16\pi G}\int d^{4}x \sqrt{-g}f(R) + S_{m}, \label{f_R_action} \ee
where $G$ is the Newtonian constant, and $S_{m}$ is the matter term in the action \cite{Carroll_2, Starobinsky, Hu_Sawicki, Miranda}. [See Refs.~\cite{Sotiriou,Tsujikawa1} for reviews of $f(R)$ theory.]

Any modified gravity model should fit the conventional standard cosmology as well as explain the current cosmic speed-up issue. Specifically, in a viable model, the Universe should have had a matter-domination epoch in the early Universe to enable the formation of large-scale structures, and it should have transited from a matter-domination epoch into the current dark-energy-domination one. Moreover, in order to be able to drive the cosmic speed-up, the effective dark energy should have sufficiently large negative pressure, and the effective equation of state should be less than $-1/3$.

The cosmological dynamics in modified gravity was analyzed in Refs.~\cite{Boyanovsky,Felder,Faraoni,Vikman,Wei,Bean,Amendola,Souza,Pogosian,Evans,Clifton,Copeland,Giambo,Abdelwahab,Ivanov,Jaime}.
The conditions for a viable matter-domination epoch and late-time acceleration were derived via an analysis in phase space in Ref.~\cite{Amendola}. In fact, the dynamics of $f(R)$ gravity closely depends on a potential defined by
$V'(\phi)\equiv dV/d\phi=(2f-\phi R)/3$, where $\phi \equiv f' = df(R)/dR$. In this paper, the conditions of cosmological viability are studied directly by considering how this potential determines the dynamics of $f(R)$ cosmology.

The currently observed value of the cosmological constant presents a hierarchy problem between the cosmological acceleration scale and the Planck scale \cite{Weinberg, Peebles}. An $f(R)$ model whose modification term has an $R\ln R$ form can generate a large hierarchy between these two scales. The model can be obtained from the arguments of the running of the gravitational coupling. In high energy physics, the renormalized coupling parameters run with the energy scale. Noting that the curvature scalar is a basic quantity in gravity that describes interaction scales, and assuming that the (classical) gravitational coupling varies with the curvature scale, one is led to $f(R)$ gravity~\cite{Frolov}. If the running is defined by a quadratic beta function, which is similar to the one in quantum chromodynamics, one ultimately obtains an $R\ln R$ model. This model does not have the singularity problem discussed in Ref.~\cite{Frolov1}. On the other hand, in this model, general relativity is recovered only for some period of curvature scale due to the logarithmic running of $f'$ with respect to the matter density. As a result, it is hard for the $R\ln R$ model to have a cosmological evolution consistent with the observations. This problem becomes less severe in a modified logarithmic model obtained by shifting the fixed point $\alpha_{*}$ of the beta function from zero to a positive value. This new model keeps the hierarchy feature; its corresponding cosmological evolution is more compatible with the observations than that in the $R\ln R$ model, but is still not ideal. The Lagrangian density of a viable $f(R)$ model (e.g., the Hu-Sawicki model) should be very close to that of the $\Lambda$CDM model so as to fit the cosmological observations in both the early and the late Universe. In this paper, we study the phase-space dynamics and the cosmological evolution of the $R\ln R$ model and the Hu-Sawicki model with the following techniques: compactifying the infinite phase space into a finite space via the $\text{Poincar\'e}$ transformation; studying the vector fields on two-dimensional slices of the constraint surface when the constraint surface is three dimensional; and plotting typical trajectories of the phase flows.

The paper is organized as follows.
In Sec.~\ref{sec:f_R_cosmology}, we construct the dynamical system for $f(R)$ cosmology.
In Sec.~\ref{sec:conditions_viability}, the conditions of the cosmological viability for $f(R)$ gravity are explored.
Secction~\ref{sec:introduction_RlnR} introduces the $R \ln R$ model.
In Secs.~\ref{sec:phase_RlnR} and \ref{sec:evolution_RlnR}, the phase-space dynamics and the cosmological evolution of the $R \ln R$ model are studied, respectively.
In Sec.~\ref{sec:dynamics_Hu_Sawicki}, we explore the phase-space dynamics of the Hu-Sawicki model.
Lastly, Sec.~\ref{sec:conclusions} summarizes our results.
%
%%%%%%%%%%%%%%%%%%%%%%%%%%%%%%%%%%%%%%%%%%%%%%%%%%%%%%%%%%%%%%%%%%%%%%%%%%%%%%%%%%%%%%%%%%%%%%%%%%%%%%%%%%%%%%%%%%%%%%%%%%
\section{The dynamical system in $f(R)$ cosmology \label{sec:f_R_cosmology}}
%%%%%%%%%%%%%%%%%%%%%%%%%%%%%%%%%%%%%%%%%%%%%%%%%%%%%%%%%%%%%%%%%%%%%%%%%%%%%%%%%%%%%%%%%%%%%%%%%%%%%%%%%%%%%%%%%%%%%%%%%%
%
In this section, we prepare for the dynamical analysis of $f(R)$ cosmology. The equivalent of the Einstein equation in $f(R)$ gravity reads,
\be f'R_{\mu\nu}-\frac{1}{2}f g_{\mu\nu} -\left(\nabla_{\mu}\nabla_{\nu}-g_{\mu\nu} \Box\right) f'
= 8\pi GT_{\mu\nu},\label{gravi_eq_fR} \ee
where $f'$ denotes the derivative of the function $f$ with respect to its argument $R$, and $\Box$ is the usual
notation for the covariant D'Alembert operator $\Box\equiv\nabla_{\alpha}\nabla^{\alpha}$. Compared to general relativity, $f(R)$ gravity has one extra scalar degree of freedom, $f'$. The dynamics of this degree of freedom is determined by the trace of Eq. (\ref{gravi_eq_fR})
\vspace{-10pt}
\be \Box f'=\frac{1}{3}(2f-f'R) + \frac{8\pi G}{3}T, \label{trace_eq1}\ee
where $T$ is the trace of the stress-energy tensor $T_{\mu\nu}$. Identifying $f'$ with a scalar degree of freedom by
\be \phi\equiv\frac{df}{dR}, \label{f_prime}\ee
and defining a potential $V(\phi)$ by
\be V'(\phi)\equiv\frac{dV}{d\phi}=\frac{1}{3}(2f-\phi R), \label{v_prime} \ee
one can rewrite Eq.~(\ref{trace_eq1}) as~\cite{Sotiriou}
\be \Box \phi=V'(\phi)+\frac{8\pi G}{3}T.\label{trace_eq2}\ee
We consider the homogeneous Universe with the flat Friedmann-Robertson-Walker metric,
\be ds^{2}=-dt^{2}+a^{2}(t)d\vec{x}^{2}, \label{FRW_metric}\ee
where $a(t)$ is the scale factor. In this case, the evolution of the Universe is described by a four-dimensional dynamical system of $\{\phi,\pi,H,a\}$, where
\be \pi\equiv\dot{\phi}, \label{pi_definition}\ee
$H$ is the Hubble parameter, and the dot $(\cdot)$ denotes the derivative with respect to the coordinate time $t$. Equation (\ref{trace_eq1}) provides the equation of motion for $\pi$
\be \dot{\pi}=-3H\pi-V'(\phi)+\frac{8\pi G}{3}{\rho_m}. \label{pi_dot}\ee
The equation of motion for $H$ is
\be \dot{H}=\frac{R}{6} -2H^{2}. \label{H_dot} \ee
The definition of the Hubble parameter implies that
\be \dot{a}=aH. \label{a_dot}\ee
The system is constrained by the Friedman equation
\be H^{2}+\frac{\pi}{\phi}H+\frac{f-\phi R}{6\phi}-\frac{8\pi G}{3\phi}(\rho_{m}+\rho_{r})=0,
\label{constraint_eq} \ee
where $\rho_m$ and $\rho_r$ are the density of matter and the density of radiation, respectively.
Equations (\ref{pi_definition})-(\ref{constraint_eq}) provide a closed description of the dynamical system $\{\phi,\pi,H,a\}$.

In order to explore whether $f(R)$ gravity can account for the cosmic speed-up, it is instructive to cast the formulation of $f(R)$
gravity into a format similar to that of general relativity. We rewrite Eq.~(\ref{gravi_eq_fR}) as
\be G_{\mu \nu}=8\pi G \left( T_{\mu \nu} + T_{\mu \nu} ^{(\text{eff})} \right),
\label{field_eq4} \ee
where
\begin{align}
8\pi GT_{\mu \nu} ^{(\text{eff})}=&\frac{f-f'R}{2}g_{\mu\nu}+\left(\nabla_{\mu}\nabla_{\nu}-g_{\mu\nu} \Box\right)f' \nonumber \\
 &+ (1-f')G_{\mu\nu}.
 \label{tilde_T4}
\end{align}
$T^{\mu\nu}_{\text{(eff)}}$ is the energy-momentum tensor of the effective dark energy. It is guaranteed to be conserved,
$T^{\mu\nu}_{\text{(eff)};\nu} = 0$. Equation (\ref{tilde_T4}) reveals the definition of the equation of state for the effective dark energy
\be w_{\text{eff}}\equiv\frac{p_{\text{eff}}}{\rho_{\text{eff}}}, \label{w_eff_tilde_1}\ee
where
\be
\begin{split}
8\pi G \rho_{\text{eff}}& =3H^{2}-8\pi G(\rho_m+\rho_r)\\
& =\frac{f'R-f}{2}-3H\dot{f'} +3H^{2}(1-f'),
\end{split}
\label{rho_eff}
\ee
\vspace{-18pt}
\be
\begin{split}
8\pi G p_{\text{eff}}& =H^{2}-R/3-8\pi G p_r\\
& =\ddot{f'}+2H\dot{f'}+\frac{f-f'R}{2}+\left(H^{2}-\frac{R}{3}\right)(1-f').
\end{split}
\label{p_eff}
\ee
In order for an $f(R)$ model to account for the cosmic speed-up, $w_{\text{eff}}$ should be less than $-1/3$.
%
%%%%%%%%%%%%%%%%%%%%%%%%%%%%%%%%%%%%%%%%%%%%%%%%%%%%%%%%%%%%%%%%%%%%%%%%%%%%%%%%%%%%%%%%%%%%%%%%%%%%%%%%%%%%%%%%%%%%%%%%%
\section{Cosmological viability \label{sec:conditions_viability}}
%%%%%%%%%%%%%%%%%%%%%%%%%%%%%%%%%%%%%%%%%%%%%%%%%%%%%%%%%%%%%%%%%%%%%%%%%%%%%%%%%%%%%%%%%%%%%%%%%%%%%%%%%%%%%%%%%%%%%%%%%
%
Many $f(R)$ models have been proposed to address the current cosmic speed-up problem. It is necessary to check whether these models agree with the observations of both the early and the late Universe. In a viable $f(R)$ theory, there should be a matter-domination epoch in the early Universe such that large-scale structures could be formed. Moreover, the Universe should experience an acceleration during late time. The conditions of cosmological viability for $f(R)$ theory were discussed via dynamical analysis in phase space in Ref.~\cite{Amendola}. With this approach, one could investigate the conditions for the existence of a viable matter-domination epoch prior to a late-time acceleration, which can be expressed as
\be m(r)\approx 0^{+} \text{ and } \frac{dm}{dr}>-1, \mbox{  at } r\approx -1, \label{matter_domin}\ee
where $m\equiv f''R/f'$ and $r\equiv -f'R/f$. Actually, $r$ and $m$ are closely related to $V'(\phi)$ and $V''(\phi)$, respectively, with $V'(\phi)$ being defined by Eq.~(\ref{v_prime}) and
\be V''(\phi)=\frac{f'-f''R}{3f''}. \label{v_dub_prime}\ee
In this section, we will revisit these cosmological viability conditions by using the scalar field description for $f(R)$ gravity.

In the standard cosmology based on general relativity, a matter-domination epoch (and also a radiation-domination epoch) is ensured in the early Universe. Therefore, in order to obtain a long matter-domination epoch in $f(R)$ gravity, one may consider how $f(R)$ gravity could be reduced to general relativity. The restoration of general relativity implies that
\be f(R) \approx  f'R, \mbox{ with } \phi\equiv f' \approx 1, \label{f_R_to_GR}\ee
which results in $r\approx -1$ shown in Eq.~(\ref{matter_domin}). In the early Universe, the matter-domination epoch should last long enough to ensure
large-scale structure formation. This means that general relativity should be restored for a long time. Therefore, $f'$ should roll down very slowly. Combining Eqs.~(\ref{pi_definition}) and (\ref{pi_dot}), one obtains
\be \ddot{\phi}=-3H\dot{\phi}-V'(\phi)+\frac{8\pi G}{3}{\rho_m}. \label{phi_evolution}\ee
Consequently, when the field $\phi$ evolves slowly, we have
\be |3H\dot{\phi}|\ll V'(\phi)\approx \frac{8\pi G}{3}{\rho_m}. \label{v_rho_dynamical} \ee
Note that $\rho_m=\rho_{m0}/a^3$ and $\dot{a}=aH$, where $\rho_{m0}$ is the matter density of the current Universe. Taking the time derivative of $V'(\phi)\approx 8\pi G\rho_m/3$, we arrive at
\be V''\cdot \dot{\phi}
\approx -8\pi G\frac{\rho_{m0}}{a^4}\dot{a}
=-8\pi G \rho_m H
\approx-3HV', \nonumber
\ee
and therefore
\be \dot{\phi}\approx-3H\frac{V'}{V''}. \label{phi_dot}\ee
Substituting Eq.~(\ref{phi_dot}) into Eq.~(\ref{v_rho_dynamical}) yields

\be |3\dot{\phi}H|\approx9H^{2}\frac{V'}{V''}\ll V'\approx \frac{8\pi G}{3}{\rho_m}.\ee
Then we have
\be V''\gg 9H^{2}\approx 3\cdot 8\pi G\rho_m. \label{v_dub_prime_matter_domi}\ee
The condition expressed by Eq.~(\ref{v_dub_prime_matter_domi}) can be interpreted as follows. Note that the potential $V(\phi)$ should have a minimum so that
there could be a dark-energy-domination epoch in the late Universe. In the early Universe, the field $\phi$ evolves slowly, and stays at the quasistatic equilibrium of $V'(\phi)\approx 8\pi G\rho_m/3$ as shown in Eq.~(\ref{v_rho_dynamical}). Thus, the field $\phi$ and the matter density $\rho_m$ are coupled. From this coupling, the field $\phi$ acquires mass. When the mass of $\phi$ is heavy [large $V''(\phi)$], it is hard to excite $\phi$. Then, the field $\phi$ stays near $1$ for a long time. Consequently, general relativity is restored for a long time and the Universe has a long matter-domination epoch. The matter density decreases slowly. The field $\phi$ then becomes light, and is eventually released from the coupling to the matter density and approaches the de Sitter minimum of the potential $V(\phi)$. Note that we only consider the case in which the potential $V(\phi)$ has a de Sitter minimum, like the case plotted in Fig.~\ref{fig:potential_RlnR}. Correspondingly, the Universe transits from the matter-domination epoch into the dark-energy-domination epoch.

Substituting Eq.~(\ref{v_dub_prime}) into Eq.~(\ref{v_dub_prime_matter_domi}), and noting that in the general relativistic limit $R\approx 8\pi G\rho_{m}$, one obtains
\be f'\gg f''R. \label{condition_GR_recovering}\ee
The condition for the recovery of general relativity is given by Eq. (\ref{v_dub_prime_matter_domi}) or Eq.~(\ref{condition_GR_recovering}).
Equation~(\ref{condition_GR_recovering}) is equivalent to $m(r)\approx 0^{+}$, shown in Eq.~(\ref{matter_domin}). Equation~(\ref{condition_GR_recovering}) can be interpreted via a comparison of the modification term and the main term of the function $f(R)$. We write the function $f(R)$ as $f(R)=R+A(R)$, where $R$ is the main term and $A(R)$ is the modification term. If $f(R)$ theory satisfies Eq.~(\ref{f_R_to_GR}) at a certain time in the early Universe, which means that $|A(R)|\ll R$ and $|A'(R)|\ll 1$, there is a matter-domination epoch at that time. However, to make this matter domination and/or the general relativity recovery last long enough, $A'(R)$ should also change with respect to $R$ more slowly than $1/R$, namely $A''(R)\ll 1/R$, as implied in Eq.~(\ref{condition_GR_recovering}).

The process of the field $\phi$ obtaining mass from its coupling to the matter density is very similar to the chameleon mechanism studied in the context of the Solar System tests of $f(R)$ gravity~\cite{Hu_Sawicki,Justin1,Justin2,Navarro,Faulkner,Gu,Tsujikawa2,Tsujikawa3,Guo}. In the chameleon mechanism, the field $\phi$ is coupled to the matter densities of the Sun and of the background, respectively. The field $\phi$ acquires a large mass from this coupling; thus, $f(R)$ gravity could in principle evade the Solar System tests.

In addition to having a long matter-domination epoch in the early Universe, a viable $f(R)$ model should also have a stable dark-energy-domination epoch in the late Universe to account for the cosmic acceleration. (The potential $V(\phi)$ needs to have a minimum.) Generally, the parameters in viable $f(R)$ models need to take values that can make a trade-off between the two requirements.
%%%%%%%%%%%%%%%%%%%%%%%%%%%%%%%%%%%%%%%%%%%%%%%%%%%%%%%%%%%%%%%%%%%%%%%%%%%%
\section{Introduction to the $R\ln R$ model\label{sec:introduction_RlnR}}
%%%%%%%%%%%%%%%%%%%%%%%%%%%%%%%%%%%%%%%%%%%%%%%%%%%%%%%%%%%%%%
Next we explore the cosmological dynamics of an $f(R)$ model, in which the modification term is described by an $R\ln R$ form. First, we briefly explain the idea of the running gravitational coupling proposed in our previous work~\cite{Frolov}. Then, we introduce the $R\ln R$ model generated by the running.

Observations of the accelerating expansion of the Universe indicate the presence of a tiny but nonvanishing cosmological constant. This implies a hierarchy problem between the (ultraviolet) Planck scale and the (infrared) cosmological acceleration scale \cite{Weinberg, Peebles}. Here, we look for possible solutions to this problem in the context of effective field theory.

In high energy physics, the renormalized coupling parameters run as beta functions of the energy scale. In gravity, the basic scale is set by the curvature of the spacetime. Assuming that the (classical) gravitational coupling varies with the curvature scalar $R$, one is led to $f(R)$ gravity. We give a brief review of this approach below.

Considering that Newton's gravitational constant runs with the Ricci scalar $R$, we introduce a dimensionless coupling $\alpha$,
\be 8\pi G=\alpha m_{\text{pl}}^{-2}, \label{alpha_definition}\ee
where $m_{\text{pl}}$ is the Planck mass. If the renormalization group flow is autonomous, the running of the dimensionless coupling $\alpha$ can be described by a beta function,
\be \mu \frac{d\alpha}{d\mu}=\beta(\alpha), \label{beta_function}\ee
where $\mu\equiv R/R_0$ and $R_0$ is a positive constant parameter. The integration of the above equation yields an $\alpha$ as a function of the curvature. Then, by replacing $8\pi G$ in the Lagrangian density of general relativity ${\cal L}_{\text{GR}}=R/(16\pi G)$ with $\alpha m_{\text{pl}}^{-2}$ [refer to Eq.~(\ref{alpha_definition})], one obtains $f(R)$ gravity with the Lagrangian density
\be
{\cal L}_{f(R)}=\frac{m_{\text{pl}}^{2}}{2}\, \frac{R}{\alpha}.
\ee
With these arguments, the power-law corrections to the Einstein-Hilbert action
\be f(R)=R\left[1+\lambda\left(\frac{R}{R_0}\right)^{n}\right]\ee
can be generated by an autonomous flow,
\be \beta(\alpha)=n\alpha(\alpha-1), \label{beta_function_power_law}\ee
with
\be \alpha\equiv\frac{R}{f(R)}=\frac{1}{1+\lambda \mu^{n}}. \label{alpha_power_law}\ee
Combining Eqs.~(\ref{beta_function}), (\ref{beta_function_power_law}), and (\ref{alpha_power_law}), one obtains the description of $\beta$ as a function of the scale $\mu$,
\be \beta=-\frac{n\lambda \mu^{n}}{(1+\lambda \mu^{n})^{2}}.\ee
At high-curvature scales, where $\mu\gg 1$, we have
\be \beta\approx -\frac{n}{\lambda \mu^{n}}.\ee
Therefore, the separation of the orders of magnitude for the beta function is comparable to that for curvature. In other words, in the power-law $f(R)$ models, a big gap in the beta function corresponds to a big gap in curvature between the Planck scale and the local environment on Earth. This is also true for some other $f(R)$ models (e.g., the Hu-Sawicki model~\cite{Hu_Sawicki}). However, a reasonable gap in the beta function can make a big hierarchy of curvature when the linear term in the beta function is absent. Consider a quadratic beta function,
\be \beta=-\alpha^{2},\ee
which is similar to the one in quantum chromodynamics. This is an \emph{ad-hoc} choice, but it leads to a model with interesting features. It generates a model with
\be f(R)=\frac{R}{\alpha_0}\left(1+\alpha_0\ln\frac{R}{R_0}\right), \label{f_R_logR_0} \ee
where $\alpha_0$ is a dimensionless constant parameter. In this model, the coupling constant runs as
\be \alpha=\frac{\alpha_0}{1+\alpha_0\ln \frac{R}{R_0}}. \ee
Due to the logarithmic relation between $\alpha$ and $R$, a small change of the orders of magnitude for $\alpha$ can generate a large hierarchy for $R$. As discussed below, this property can be used to generate small numbers to address the hierarchy problem, which is related to the big gap between the Planck scale and the cosmological constant scale.

For ease of operation, one can absorb the constant $\alpha_0$ in the denominator of Eq.~(\ref{f_R_logR_0}) into the definition of the Planck mass $m_{\text{pl}}$, and rewrite Eq.~(\ref{f_R_logR_0}) as
\be f(R)=R\left(1+\alpha_0\ln\frac{R}{R_{0}}\right). \label{f_R_logR}\ee
Therefore, in this model the scalar degree of freedom is
\be \phi \equiv f'=1+\alpha_0+\alpha_0 \ln \frac{R}{R_0}, \label{f_prime_RlnR}\ee
and the potential is determined by
\be V'(\phi)=\frac{1}{3}\Lambda e^{\phi/\alpha_0-2} (\phi-2\alpha_0). \label{v_prime_RlnR}\ee
Then, at the de Sitter point where $V'(\phi)=0$, $\phi$ is equal to $2\alpha_0$. The corresponding curvature, which is usually called the de Sitter curvature,
\be \Lambda\equiv R_{0}e^{-1/\alpha_{0}+1},\label{dS_curvature_RlnR}\ee
is exponentially suppressed compared to $R_0$. $f'$ has to be positive to avoid ghosts~\cite{Nunez} and $f''$ has to be positive to avoid the Dolgov-Kawasaki instability~\cite{Dolgov}. For the $R\ln R$ model, given Eqs.~(\ref{f_prime_RlnR}) and (\ref{dS_curvature_RlnR}), the first requirement that $f'$ be positive can be satisfied as long as the Ricci scalar is not too much smaller than the de Sitter curvature. For the $R\ln R$ model, $f''$ is equal to $\alpha_0/R$ and $\alpha_0$ is a positive constant, and we only consider the positive Ricci scalar; then, the second requirement of $f''$ being positive can be easily met.

\begin{figure}%[!htbp]
\includegraphics[width=7.5 cm]{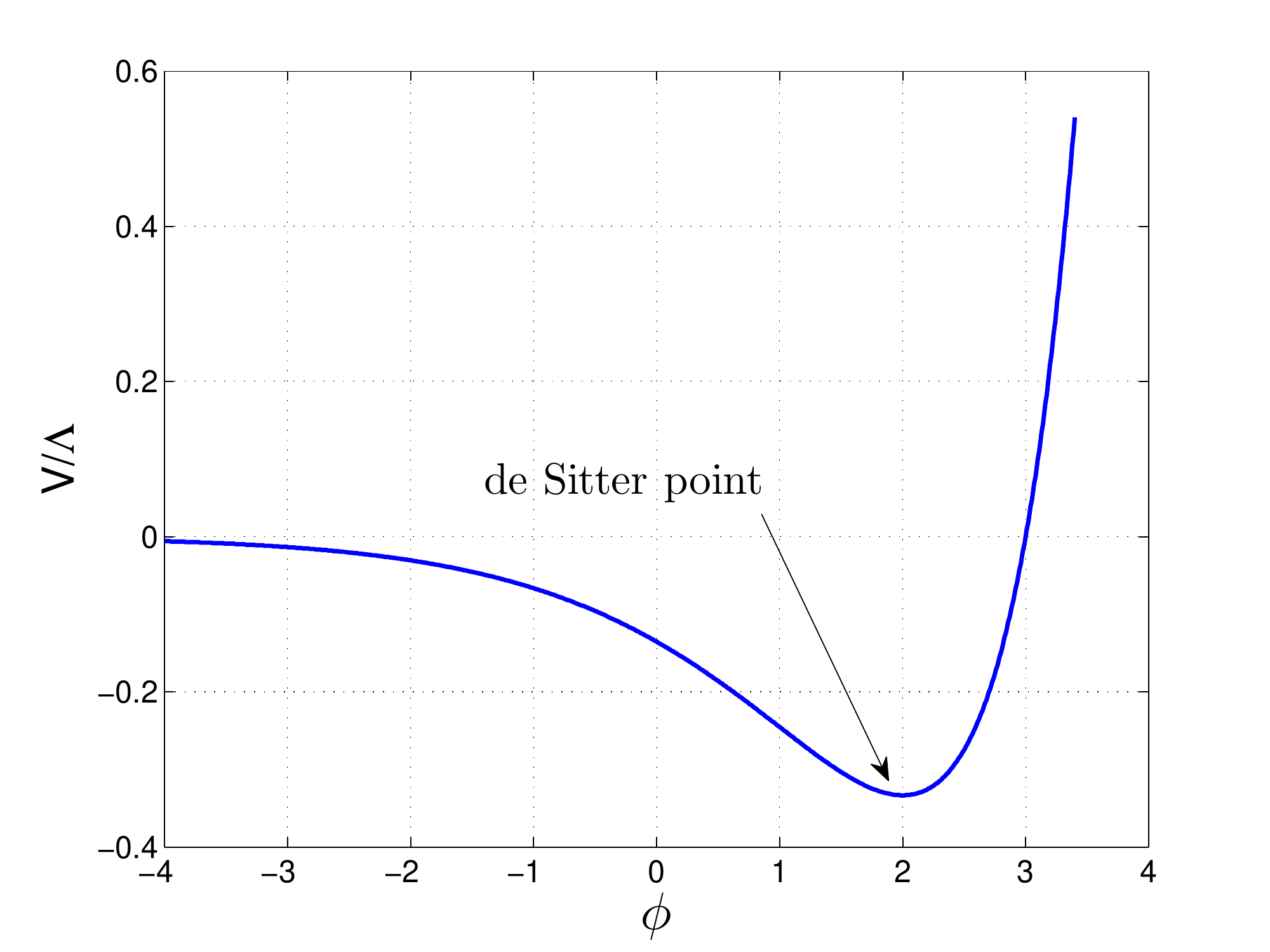}
\caption{The potential $V(\phi)$ as in Eq.~(\ref{potential_RlnR}) for the $R\ln R$ model with $\alpha_0=R_0=1$.} \label{fig:potential_RlnR}
\end{figure}

For some $f(R)$ models, such as the Starobinsky model~\cite{Starobinsky}, $\phi$ asymptotes to a constant as the Ricci scalar goes to infinity, and the height of the potential barrier is finite. Therefore, the force from the matter density can easily push the field $\phi$ to the barrier of $V(\phi)$, and then the Ricci scalar becomes singular~\cite{Frolov1}. However, the $R\ln R$ model is free of this singularity problem. Integrating Eq.~(\ref{v_prime_RlnR}), we obtain the potential
\be V(\phi)=\frac{1}{3}\alpha_0\Lambda e^{\frac{\phi}{\alpha_0}-2}(\phi-3\alpha_0), \label{potential_RlnR}\ee
which has an exponential wall, avoiding the singularity problem. The potential is shown in Fig.~\ref{fig:potential_RlnR}.

For this model, with Eq.~(\ref{f_prime_RlnR}), the function $f(R)$ expressed by Eq.~(\ref{f_R_logR}) can be rewritten as
\be f(R)=R(\phi-\alpha_0). \label{f_R_logR_2}\ee
When the $R\ln R$ gravity is reduced to general relativity, $\phi$ evolves slowly. Then, from Eq.~(\ref{pi_dot}), which describes the dynamics of $\phi$, one obtains
\be \phi \approx 2\alpha_0 + \alpha_0 W(X), \label{phi_RlnR} \ee
where $X=8\pi G\rho/\Lambda$ and $W(X)$ is the Lambert $W$ function. The basic properties of $W(X)$ with positive $X$ are described in the Appendix.
Equations (\ref{f_R_logR_2}) and (\ref{phi_RlnR}) show that when general relativity is restored, we have
\be \alpha_0 \approx \frac{1}{W(X)} \ll 1. \nonumber \ee
When $X$ is much greater than 1, $W(X)$ is approximately equal to $\ln (X)$, as discussed in the Appendix. This feature, together with Eq.~(\ref{phi_RlnR}), implies that the field $\phi$ logarithmically runs depending on $X$, when $X$ is much greater than $1$. Therefore, this model is reduced to general relativity only for a certain period of curvature or matter density. The smaller $\alpha_0$ is, the longer the general relativity restoration period is. This is quite different from some other models, such as the Hu-Sawicki model~\cite{Hu_Sawicki} and the Starobinsky model~\cite{Starobinsky}, in which $f(R)$ gravity goes to general relativity once $\rho_m$ is above the cosmological constant scale.
On the other hand, for the $R\ln R$ model to have a sensible cosmic acceleration in the late Universe, the de Sitter curvature--and hence $\alpha_0$--cannot be too small [see Eq.~(\ref{dS_curvature_RlnR})]. Consequently, an appropriate value for $\alpha_0$ needs to be chosen to reconcile the tension between the requirements in the early and the late Universe.

%%%%%%%%%%%%%%%%%%%%%%%%%%%%%%%%%%%%%%%%%%%%%%%%
\section{Phase-space dynamics of the $R\ln R$ model \label{sec:phase_RlnR}}
%%%%%%%%%%%%%%%%%%%%%%%%%%%%%%%%%%%%%%%%%%%%%%%%
%
In this section, we study the cosmic dynamics of the $R\ln R$ model in phase space. The cosmic dynamics of $f(R)$ gravity is described by Eqs.~(\ref{pi_definition})-(\ref{constraint_eq}), shown in Sec.~\ref{sec:f_R_cosmology}.
For the $R\ln R$ model, the equations of motion (\ref{pi_definition})-(\ref{a_dot}) can be rewritten as
\be \pi\equiv\dot{\phi}, \label{pi_definition_RlnR}\ee
\vspace{-18pt}
\be \dot{\pi}=-3H\pi-\frac{1}{3}\Lambda e^{\frac{\phi}{\alpha_0}-2} (\phi-2\alpha_0)+\frac{8\pi G}{3}{\rho_m}, \label{pi_dot_RlnR}\ee
\vspace{-18pt}
\be \dot{H}=\frac{1}{6}\Lambda e^{\frac{\phi}{\alpha_0}-2} -2H^{2}, \label{H_dot_RlnR} \ee
\vspace{-18pt}
\be \dot{a}=aH. \label{a_dot_RlnR}\ee
The constraint equation (\ref{constraint_eq}) becomes
\be H^{2}+\frac{\pi}{\phi}H-\frac{\Lambda}{6\phi}e^{\frac{\phi}{\alpha_0}-2}-\frac{8\pi G}{3\phi}(\rho_{m}+\rho_{r})=0.
\label{constraint_eq_RlnR} \ee

\subsection{Phase-space dynamics in vacuum}
\begin{figure*}
\includegraphics[width=17 cm]{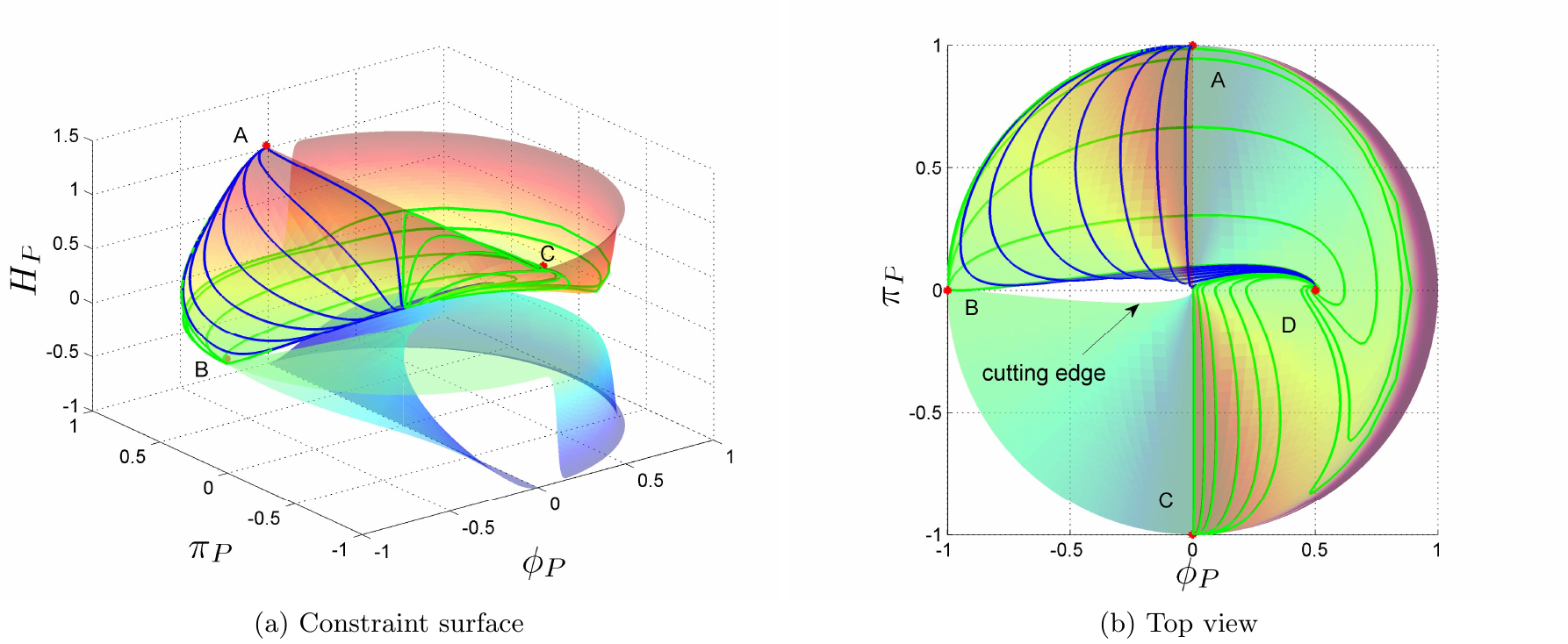}
  \caption{(Color online) The constraint surface and the phase-space flows with $\rho_m=0$ for the $R\ln R$ model with $\alpha_0=R_0=1$. The phase currents flow out of Point $A$ and move to Point $D$. Point $A$ is a repeller, Points $B$ and $C$ are saddle points, and Point $D$ is an attractor. Regarding the trajectories in green (light color), the parts of them between $A$ and $C$ are not plotted due to the difficulty in obtaining an accurate numerical integration near the boundary. The shadings correspond to the values of $H_P$. A color bar is not shown because the values of $H_P$ can be seen from the $z$ axis.}
  \label{fig:surface_and_diagram}
\end{figure*}

For simplicity, let us first consider the dynamics in vacuum, where both $\rho_m$ and $\rho_r$ are equal to zero. In this case, the solutions to the constraint
equation (\ref{constraint_eq_RlnR}) are
\be
  H_{\pm} = \frac{1}{2}\left[-\frac{\pi}{\phi}\pm \sqrt{\left(
\frac{\pi}{\phi}\right)^{2} +\frac{2\Lambda}{3\phi}e^{\frac{\phi}{\alpha_0}-2}} \right].
\label{H_plus_minus} \ee
Since the domains of definition of $\{\phi, \pi, H\}$ span from $-\infty$ to $+\infty$, it is hard to directly view the global structure of the constraint surface in the space of $\{\phi, \pi, H\}$. Instead, we use the $\text{Poincar\'e}$ compactification in the cylindrical coordinate system to transform $\phi$, $\pi$, and $H$, respectively, to
\be
\begin{array}{l l}
  \phi_P \equiv \frac{\phi}{\sqrt{\sigma+\phi^{2}+\pi^{2}}},\\
  \\
  \pi_P \equiv \frac{\pi}{\sqrt{\sigma+\phi^{2}+\pi^{2}}},\\
  \\
  H_P \equiv \frac{H}{\sqrt{\sigma+H^{2}}},
\end{array}
\label{Poincare_transform}
\ee
where $\sigma$ is an arbitrary constant, and we set it to $12$ for the $R\ln R$ model. In this way, the constraint surface is compactified into a finite space, as shown in Fig.~\ref{fig:surface_and_diagram}. The Hubble parameter in the upper branch of the constraint surface is positive, corresponding to an expanding Universe, whereas the lower branch corresponds to a contracting one. The constraint surface is folded in the octants of $(\phi\le0, \pi\ge0, H\ge0)$ and $(\phi\le0, \pi\le0, H\le0)$. On the folding line, which we also call a cutting edge, the solutions of $H_{+}$ and $H_{-}$ merge and become equal. This and Eq.~(\ref{H_plus_minus}) together imply that the cutting edge can be described by
\be \pi_{\pm}=\pm \sqrt{-\frac{2}{3}\Lambda e^{\frac{\phi}{\alpha_0}-2}\phi}. \label{edge_equation}\ee
When $\phi$ goes to $-\infty$, the $\pi_{\pm}$ approach $0^{\pm}$ and the cutting edge is almost closed, as shown in Fig.~\ref{fig:surface_and_diagram}.
We denote the two ends of the cutting edge as Point $B (\phi_{P}=-1, \pi_{P}=0^{+}, H_{P}=0^{+})$ and Point $B' (\phi_{P}=-1, \pi_{P}=0^{-}, H_{P}=0^{-})$, respectively.
The two branches of the constraint surface are disconnected. The reasons for this fact are explained below.

We consider the constraint equation (\ref{H_plus_minus}). For positive $\phi$, we have $2\Lambda\exp(\phi/\alpha_0-2)/(3\phi)>0$. Consequently, we have $H_{+}>0$ and $H_{-}<0$. For negative $\phi$, the expansion branch of the constraint surface belongs to the space of $(\phi<0, \pi>0)$, while the contraction branch belongs to the space of $(\phi<0,\pi<0)$.
As shown in Fig.~\ref{fig:surface_and_diagram}, the two branches are close to each other around Point $B$ and Point $B'$.
As implied by Eq.~(\ref{edge_equation}), at Point $B$, $\pi_P=0^{+}$, and at Point $B'$, $\pi_P=0^{-}$. Therefore, $B$ and $B'$ are separated, and then the two branches are separated as well. In summary, the two branches of the constraint surface are disconnected, although they both asymptote to the point
$(\phi_{P}=-1, \pi_{P}=0, H_{P}=0)$ when compactified.

In the vacuum evolution, the phase-space flows stay on the constraint surface. Some typical trajectories of the flows with $H_P>0$ are plotted in Fig.~\ref{fig:surface_and_diagram}. There are four special points on the branch with $H_P>0$ of the constraint surface as listed below.
\begin{equation}
\begin{aligned}
& A:(\phi_{P}=0^{-}, \pi_{P}=1, H_{P}=1);\\
& B:(\phi_{P}=-1, \pi_{P}=0^{+}, H_{P}=0^{+});\\
& C:(\phi_{P}=0^{+}, \pi_{P}=-1, H_{P}=1);\\
& D:\left(\phi_{P}=0.5, \pi_{P}=0, H_{P}=\frac{\sqrt{\Lambda/(12\alpha_0)}}{\sqrt{\sigma+\Lambda/(12\alpha_0)}}=0.083\right).
\end{aligned} \nonumber
\end{equation}
At Point $A$, the kinetic term $\pi_P$ is dominant over the field $\phi_P$, $\dot{\phi}_P=-2$, and $\dot{\pi}_P=\dot{H}_P=0$. All of the phase currents flow out of Point $A$ and move to Point $D$. Therefore, loosely speaking, Point $A$ is a repeller. Point $B$ is at one end of the cutting edge. At Point $B$, the field $\phi_P$ is dominant over the kinetic term $\pi_P$, and $\dot{\phi}_P=\dot{\pi}_P=\dot{H}_P=0$. Moreover, near Point $B$, the currents slowly approach and then move away from Point $B$. Thus, Point $B$ is a saddle point. Similarly, Point $C$ is also a saddle point. At Point $C$, the kinetic term $\pi_P$ is dominant over the field $\phi_P$.
When $\rho_{m}$ is equal to zero, Eq.~(\ref{trace_eq2}) reads
\be \ddot{\phi}+3H\dot{\phi}+V'(\phi)=0. \ee
Therefore, on  the  upper  branch of the constraint  surface  with  $H_P>0$,  due  to  the  friction  force  $-3H\dot{\phi}$,
the field $\phi$ will  eventually  arrive  and  stay  at  the minimum of  the potential, where $V'(\phi)=0$, $\phi=2\alpha_0$, $\pi=0$, and $H=\sqrt{\Lambda/(12\alpha_0)}$. This minimum corresponds to Point $D$ in Fig.~\ref{fig:surface_and_diagram}, which is an attractor and is also called a de Sitter point. When the field $\phi$ comes to this point, only dark energy exists in the Universe, with normal matter diluted away.

We project the phase diagrams onto the regular space $(\phi,\pi,H)$. Near the cutting edge, the directions of the flows are described as
\be
\left. \frac{d\pi}{d\phi}\right|_{\text{flow}} = \frac{\dot{\pi}}{\dot{\phi}} = -(\phi+1)\sqrt{-\frac{\Lambda}{6\phi}e^{\frac{\phi}{\alpha_0}-2}}. \label{slope_flow} \ee
On the other hand, with Eq.~(\ref{edge_equation}), the slope of the tangent to the edge yields the same expression. To conclude, the phase-space flows are tangential to the cutting edge and do not enter the forbidden area enclosed by the edge. In other words, the constraint equation forces the currents to stay on the surface. These conclusions also apply to the compactified space $\{\phi_P,\pi_P,H_P\}$.

The corresponding behavior of the phase currents on the lower branch of the constraint surface with $H_P<0$ can be analyzed in a similar way. There are still four critical points on this branch, as listed below.
\begin{equation}
\begin{aligned}
& A':(\phi_{P}=0^{+}, \pi_{P}=1, H_{P}=-1);\\
& B':(\phi_{P}=-1, \pi_{P}=0^{-}, H_{P}=0^{-});\\
& C':(\phi_{P}=0^{-}, \pi_{P}=-1, H_{P}=-1);\\
& D':\left(\phi_{P}=0.5, \pi_{P}=0, H_{P}=-0.083\right).
\end{aligned} \nonumber
\end{equation}
The phase flows originate from the repeller Point $D'$, and terminate at the attractor Point $C'$. Point $A'$ and Point $B'$ are saddle points.

\subsection{Phase-space dynamics in the presence of matter}
\begin{figure*}[t!]
  \hspace{-18pt}
  \begin{tabular}{ccc}
    \epsfig{file=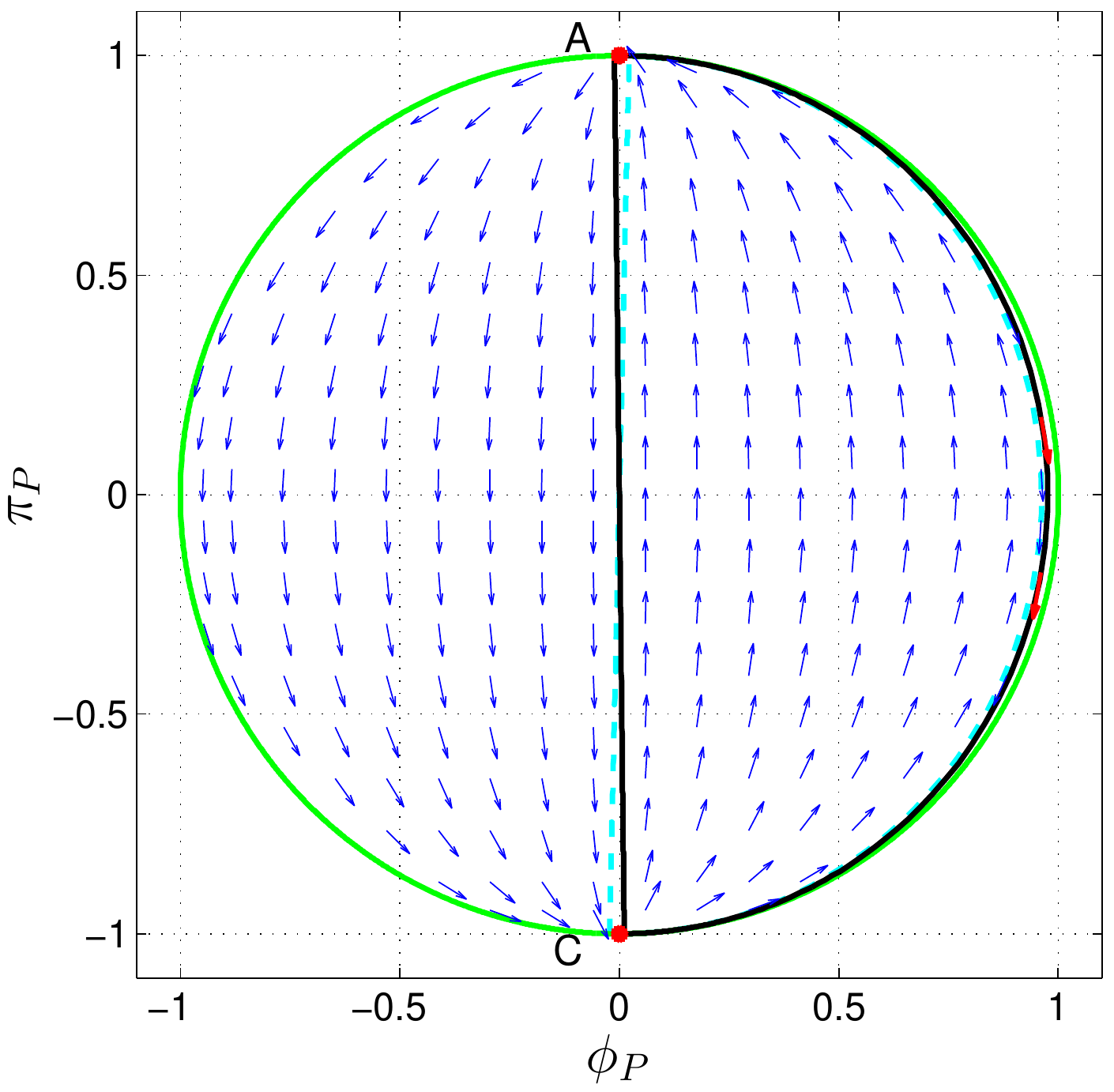, width=5.98cm} &
    \epsfig{file=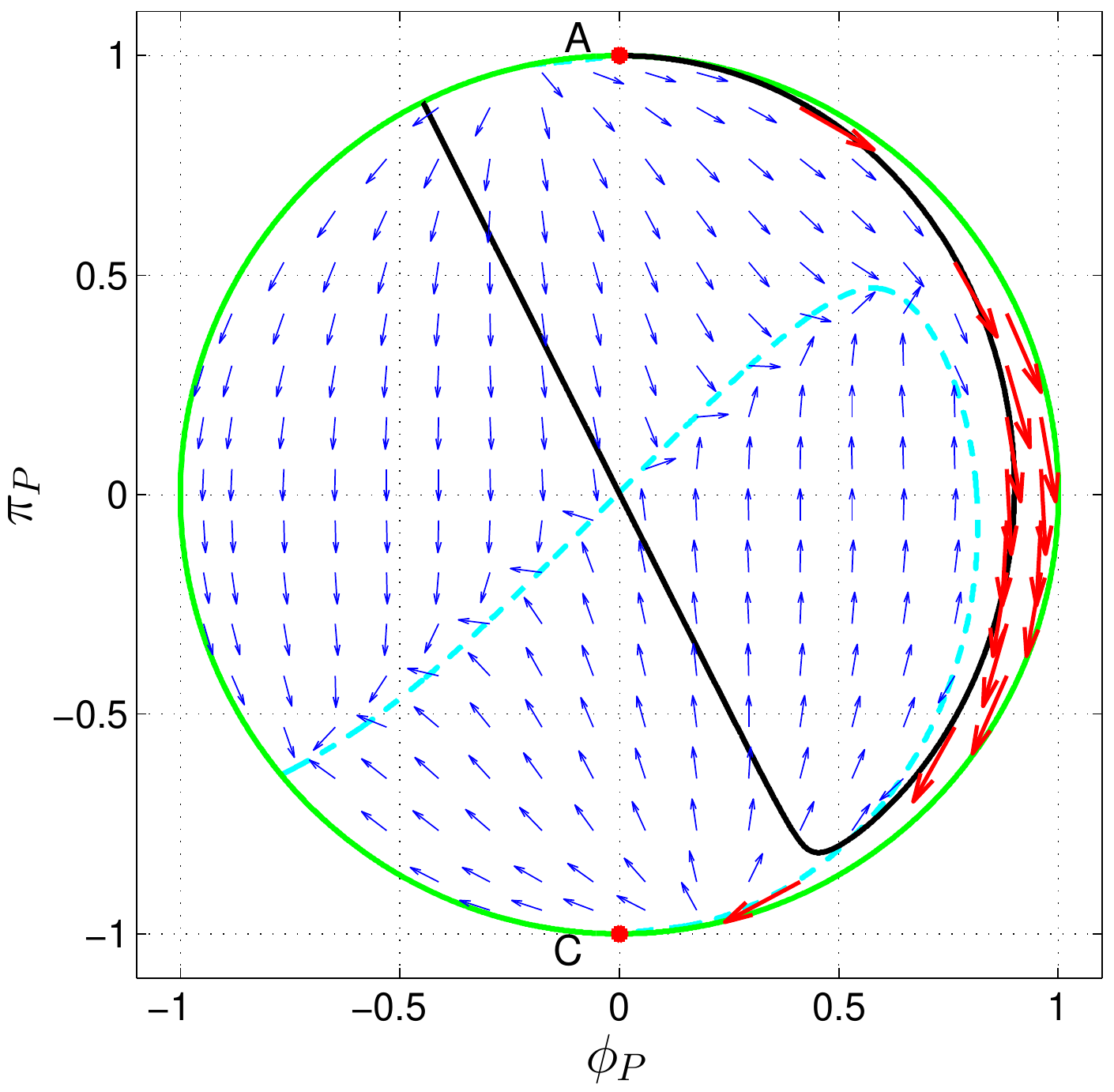, width=6cm} &
    \epsfig{file=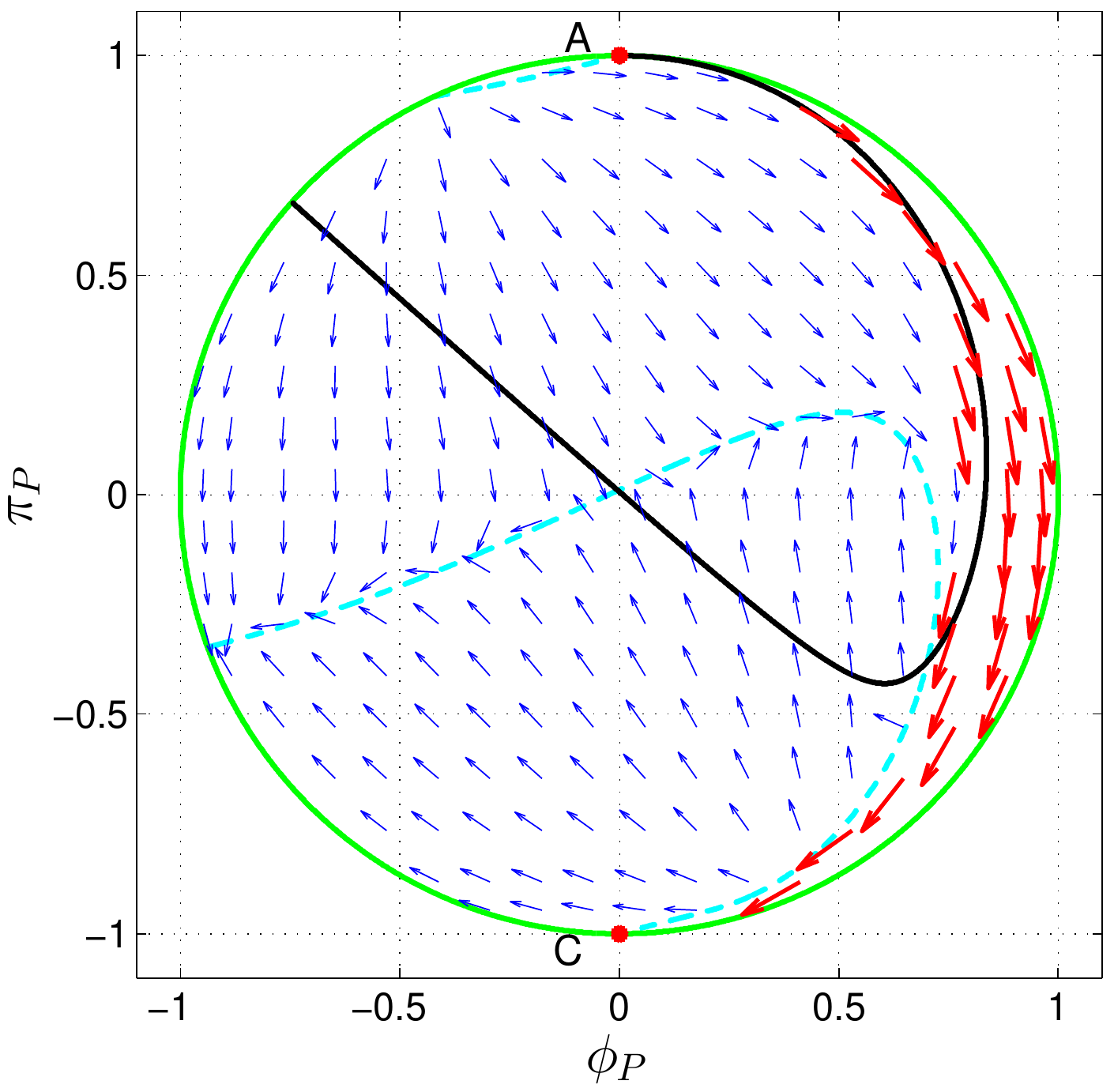, width=6cm} \\
    (a) $H_P\rightarrow1$ &
    (b) $H_P=0.5$ &
    (c) $H_P=0.25$ \\
    \epsfig{file=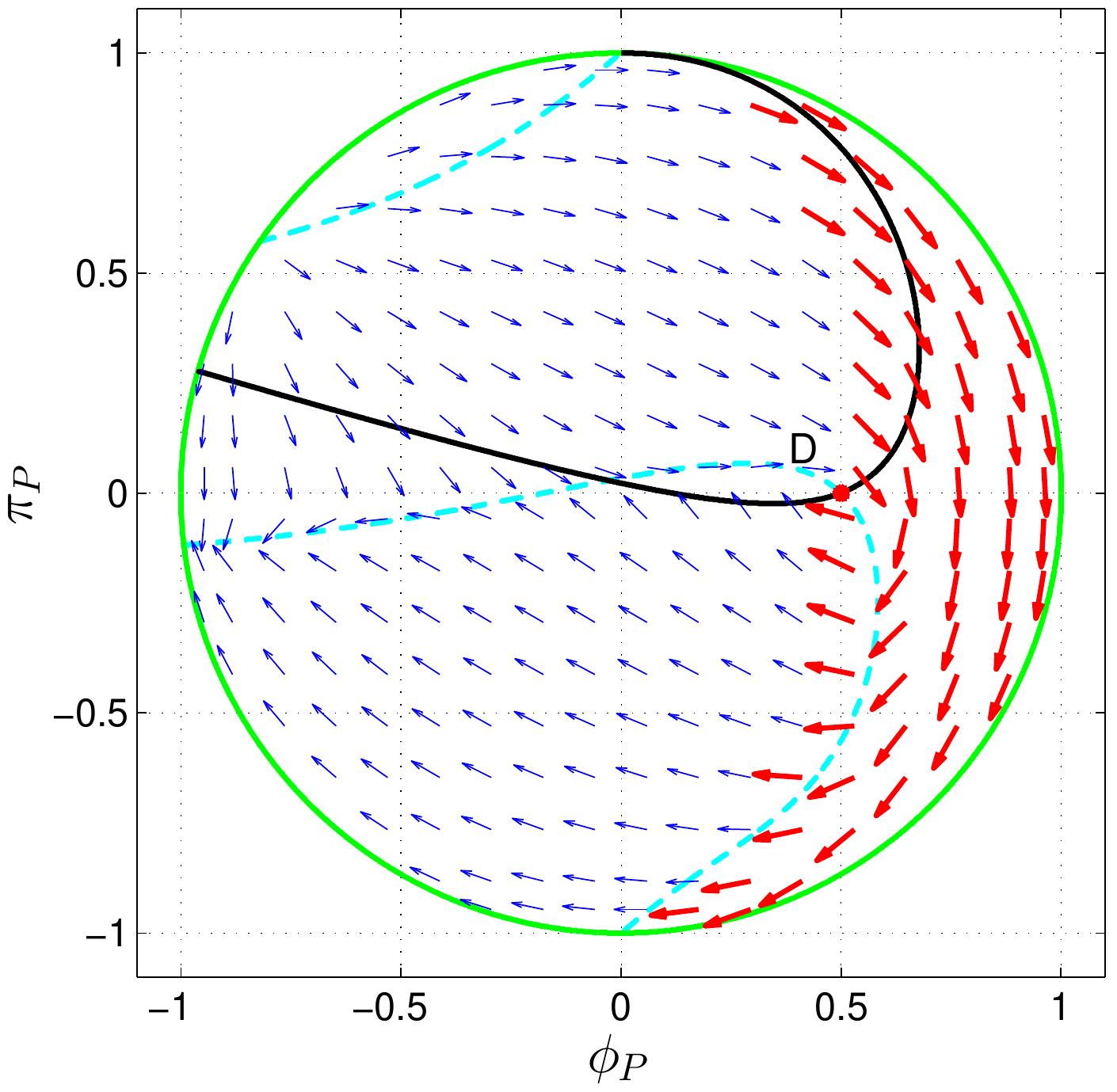, width=6cm} &
    \epsfig{file=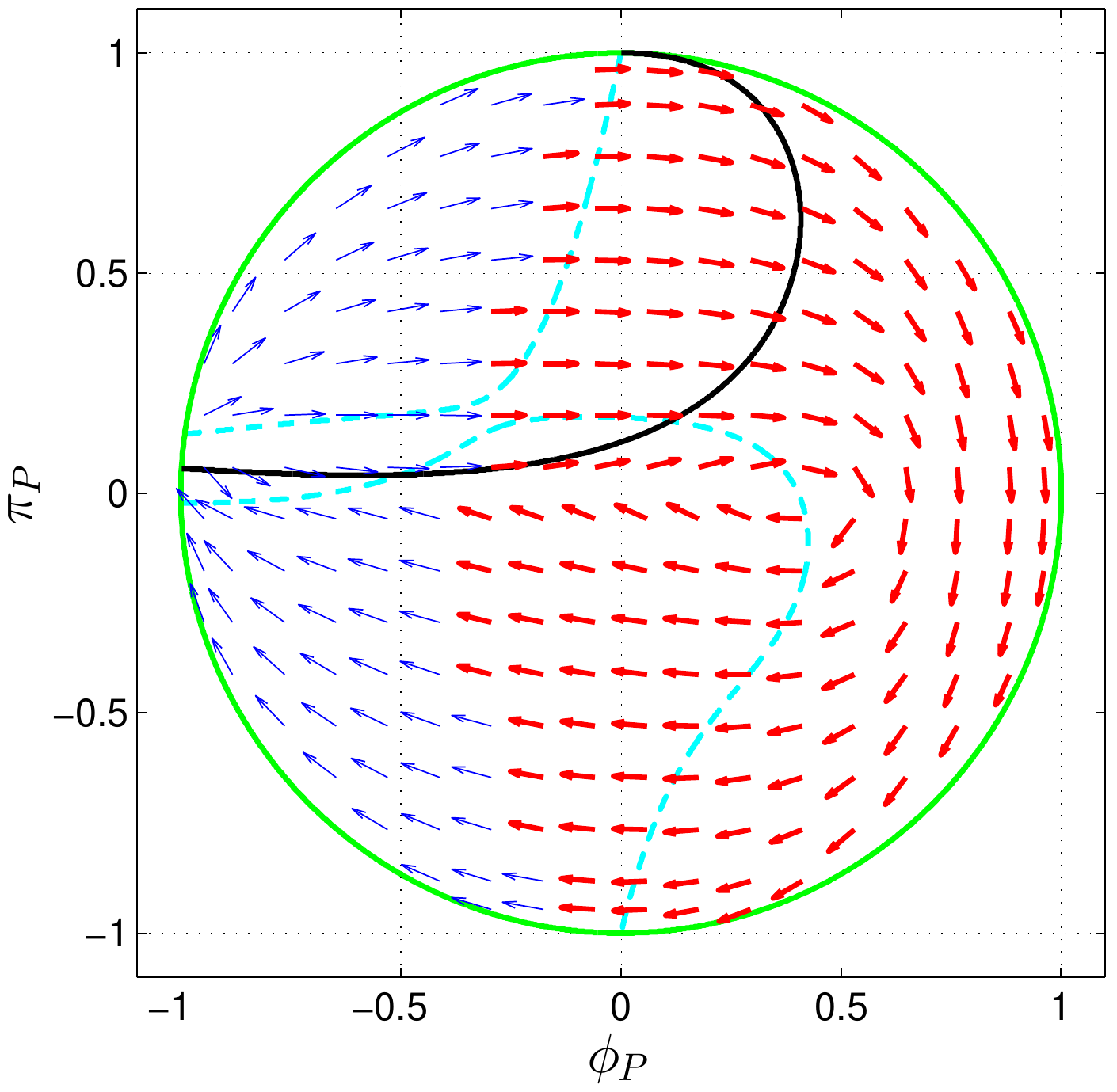, width=6cm} &
    \epsfig{file=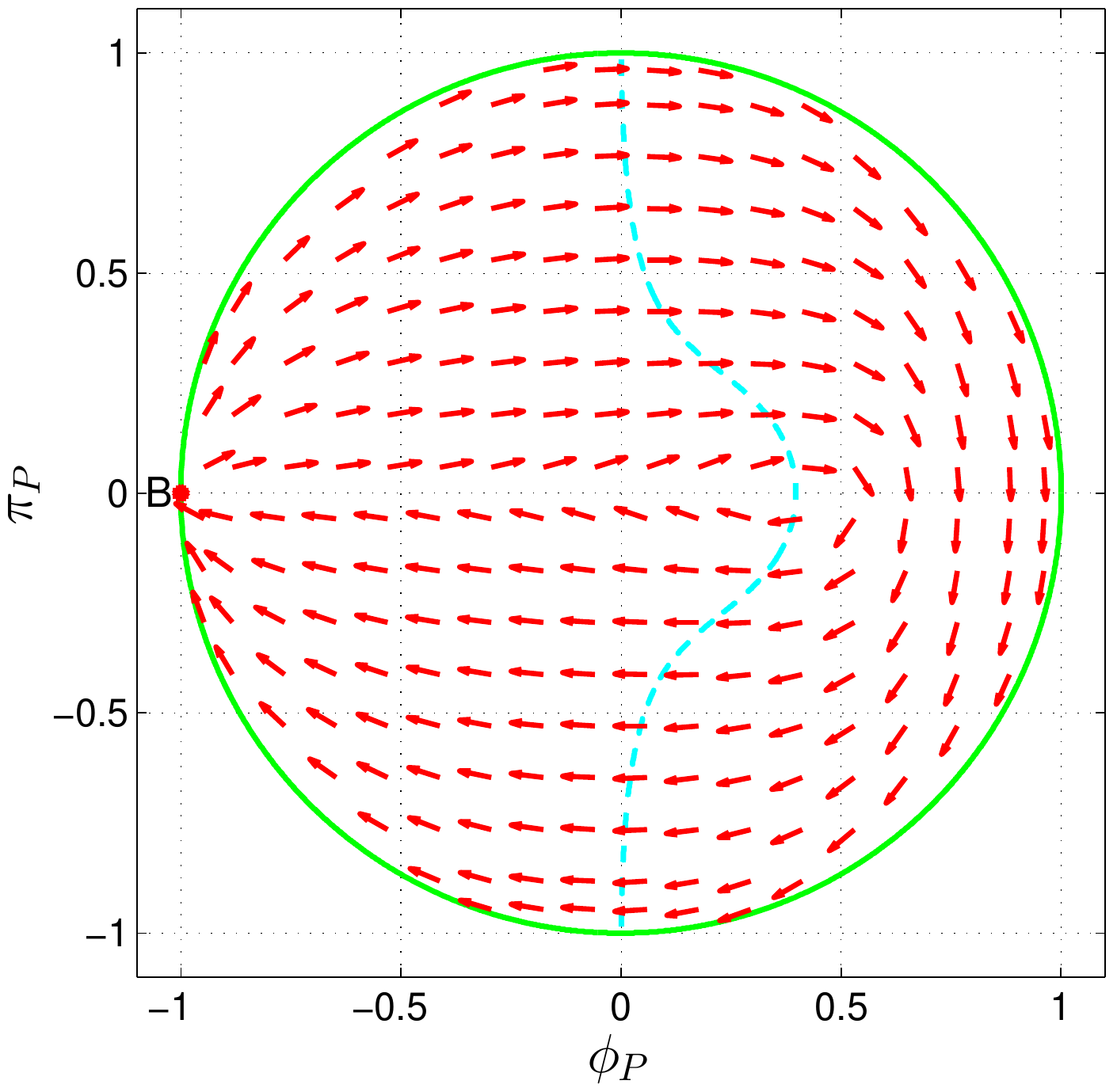, width=6cm} \\
    (d) $H_P=0.083$ &
    (e) $H_P=0.016$ &
    (f) $H_P=0$\\
    \epsfig{file=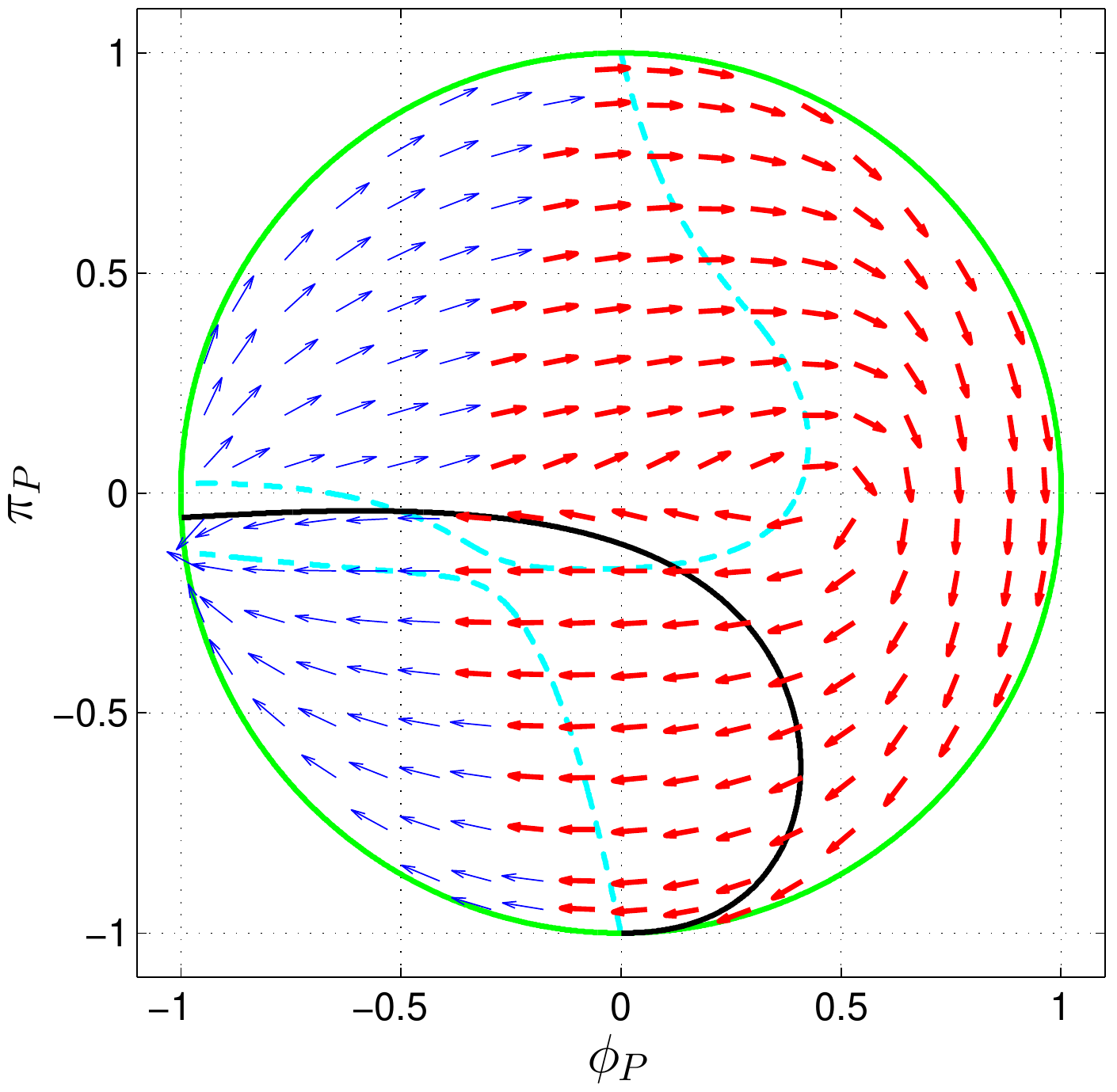, width=6cm} &
    \epsfig{file=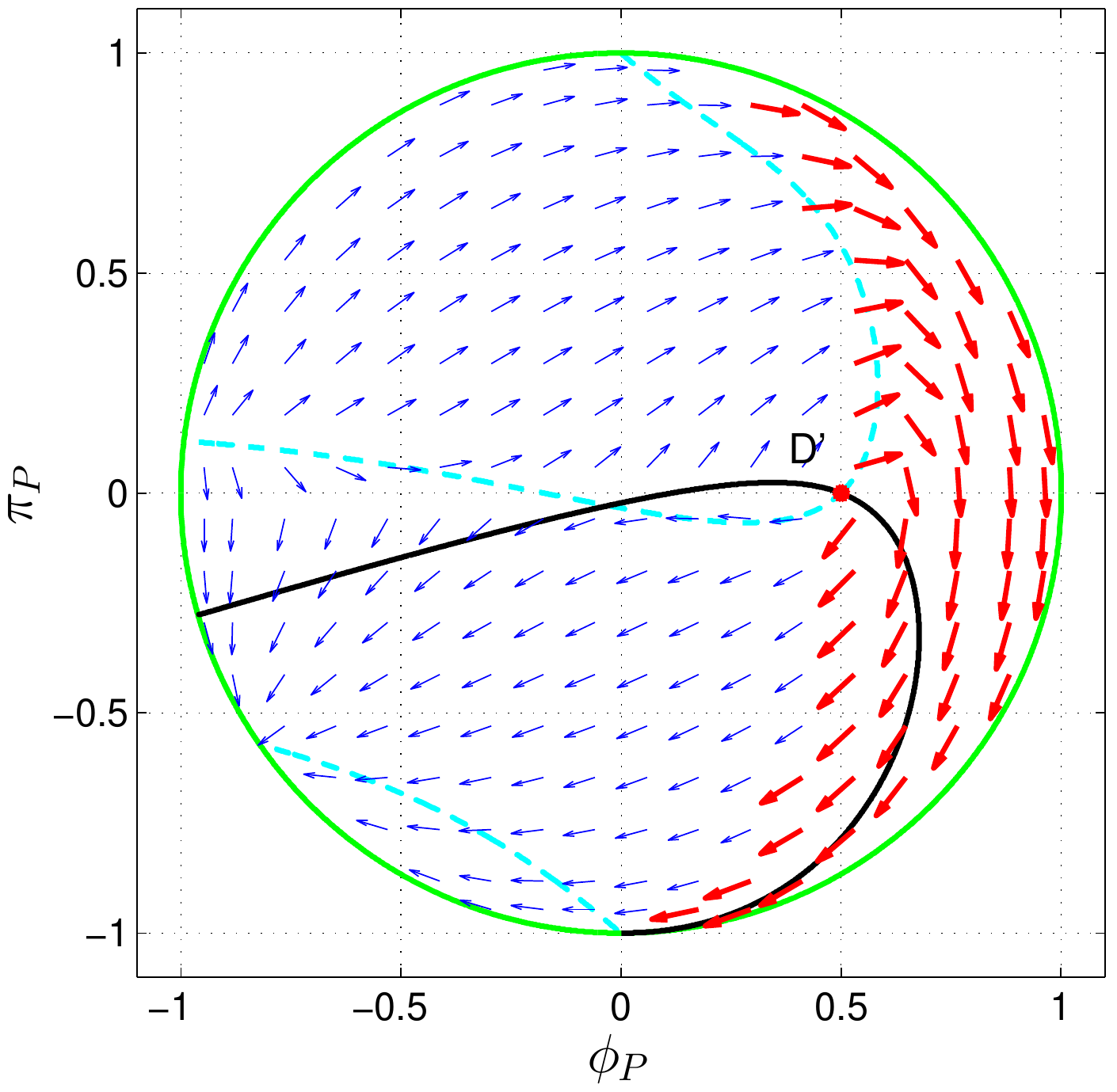, width=6cm} &
    \epsfig{file=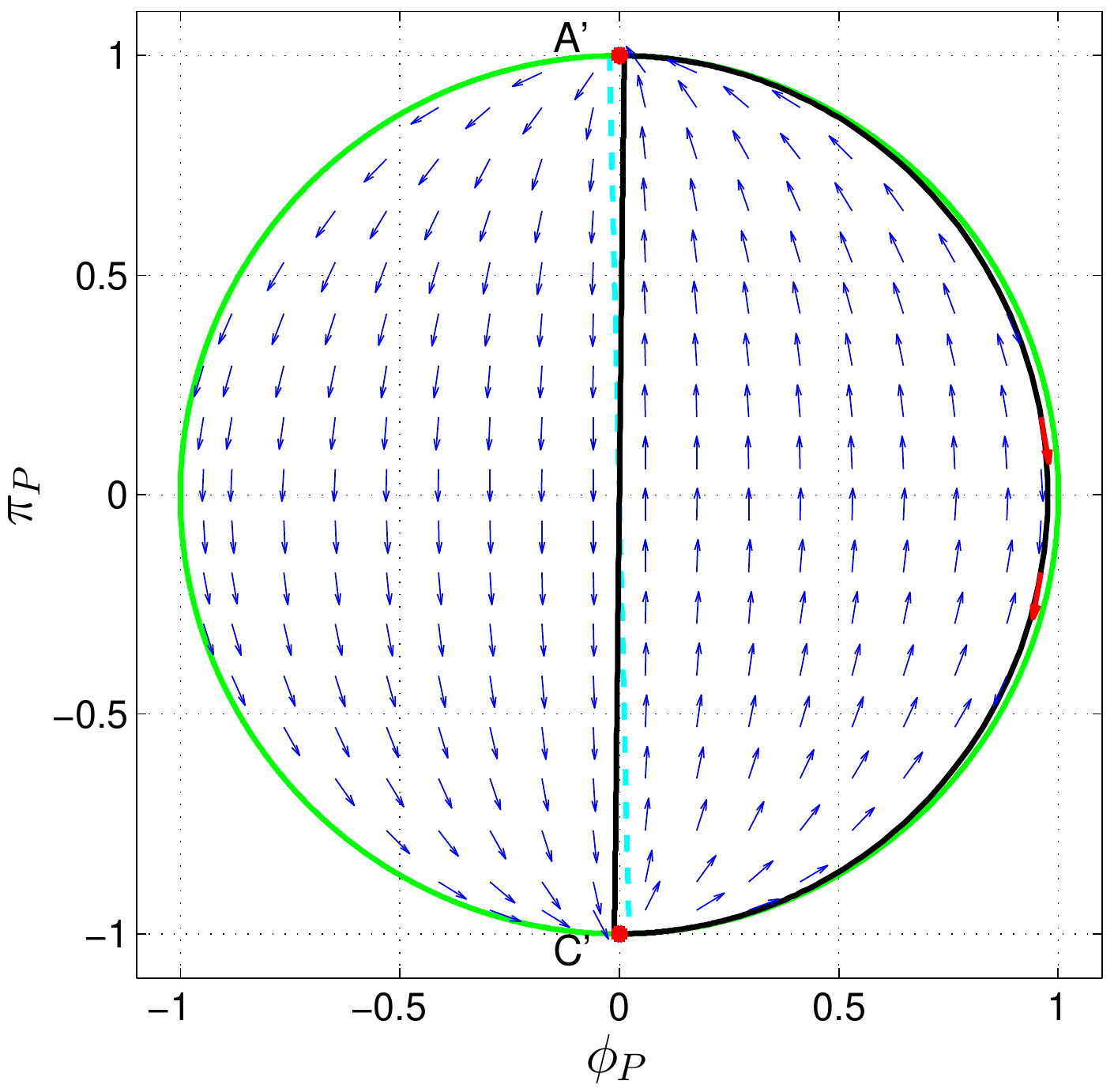, width=6cm} \\
    (g) $H_P=-0.016$ &
    (h) $H_P=-0.083$ &
    (i) $H_P\rightarrow-1$\\
  \end{tabular}
  \caption{(Color online) The vector fields of $\{\dot{\phi}_{P},\dot{\pi}_{P},\dot{H}_{P}\}$ on the slices of $H_P=\mbox{const}$ for the $R\ln R$ model with $\alpha_0=R_0=1$. The thinner (blue) arrows denote that $\dot{H_P}<0$ at the positions of the arrows. The thicker (red) arrows are for $\dot{H}_{P}>0$. The solid (black) line is the intersection between the two-dimensional vacuum constraint surface and the slice of $H_P=\text{const}$. The dashed (cyan) line is the trace of $\dot{\pi}_{P}=0$, where the flows change the direction of the $\dot{\pi}_{P}$ component.
  In (a)-(c), Point $A$ is a repeller and Point $C$ is a saddle point.
  In (d), Point $D$ is an attractor.
  In (f), Point $B$ is a saddle point.
  In (h), Point $D'$ is a repeller.
  In (i), Point $A'$ is a saddle point and Point $C'$ is an attractor.
  }
  \label{fig:vector_fields}
\end{figure*}

\begin{figure*}
\includegraphics[width=17 cm]{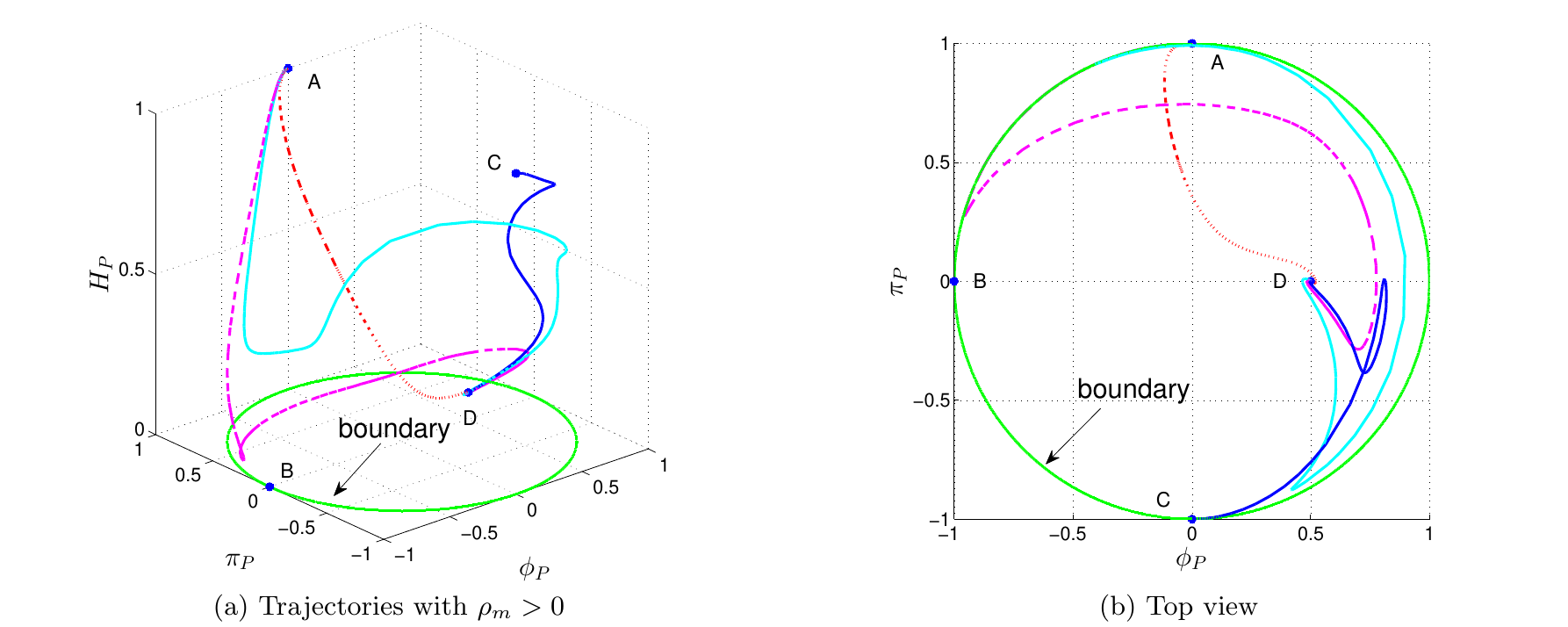}
  \caption{(Color online) Some typical trajectories of the phase-space flows with $\rho_m>0$ for the $R\ln R$ model with $\alpha_0=R_0=1$. Compared to the vacuum case, in the $\rho_m>0$ case the phase-space flows still originate at Point $A$ and terminate at Point $D$. Regarding the trajectory plotted in a solid (blue) line, one part of it between $A$ and $C$ is not shown due to the difficulty in obtaining an accurate numerical integration near the boundary.}
  \label{fig:trajectory_positive_density}
\end{figure*}

The constraint surface described by Eq.~(\ref{constraint_eq_RlnR}) is three dimensional when the matter/radiation density is not zero. For ease of visualization, we explore the vector fields of $\{\dot{\phi}_{P},\dot{\pi}_{P},\dot{H}_{P}\}$ on the slices where $H_P=\text{const}$ in the three-dimensional space $\{\phi_{P},\pi_{P},H_{P}\}$, with the scale factor $a$ being taken as an implicit variable and $\rho_r$ equal to zero.

Some typical slices of the vector fields $\{\dot{\phi}_{P},\dot{\pi}_{P}\}$ with $H_P$ taking different values from $-1$ to $1$ are shown in Fig.~\ref{fig:vector_fields}. The thinner (blue) arrows denote that $\dot{H_P}<0$ at the positions of the arrows, and the thicker (red) arrows are for $\dot{H}_{P}>0$. The solid (black) line is the intersection between the two-dimensional constraint surface of Eq.~(\ref{H_plus_minus}) and the slice of $H_P=\text{const}$. The dashed (cyan) line is the trace of $\dot{\pi}_{P}=0$, and Point $C$ is at one end of this trace. The two-dimensional constraint surface described by Eq.~(\ref{H_plus_minus}) is the separation surface for the signs of the matter density term. The matter density is positive in the space enclosed by the constraint surface, and is negative outside of the surface. The vector fields and some typical trajectories of the phase-space flows can be combined together to study the tendencies of the phase flows, as done below.

To complement the vector-field-slice approach, some typical trajectories of the phase-space flows with $\rho_m>0$ are plotted in Fig.~\ref{fig:trajectory_positive_density}. Compared to the vacuum case, the phase-space flows still originate at Point $A$ and terminate at Point $D$, but the trajectories between Point $A$ and Point $D$ can be different from the vacuum solutions.
Some flows, such as those plotted with solid (cyan) and dashed (magenta) lines in Fig.~\ref{fig:trajectory_positive_density}, behave similarly to those in the vacuum case shown in Fig.~\ref{fig:surface_and_diagram}.
The flows go down from Point $A$, then up, then make a turn and go down to Point $D$. This is also shown in Figs.~\ref{fig:vector_fields}(a)-\ref{fig:vector_fields}(c).  The thinner (blue) arrows near Point $A$ in Fig.~\ref{fig:vector_fields}(a) show the downward movement from Point $A$. The thicker (red) arrows near the boundary in Figs.~\ref{fig:vector_fields}(b) and \ref{fig:vector_fields}(c) show the upward movement. The thinner (blue) and thicker (red) arrows near the dashed (cyan) line in Figs.~\ref{fig:vector_fields}(b) and \ref{fig:vector_fields}(c) show the turn and movement down to Point $D$.
Some flows (well away from the constraint surface in vacuum), such as those plotted in dash-dotted (red) and solid (blue) lines in Fig.~\ref{fig:trajectory_positive_density}, can be very different from those in the vacuum case. The flow plotted with a dash-dotted (red) line goes directly downwards from Point $A$ to Point $D$. This behavior can also be observed from the thinner (blue) arrows in the region $-0.5<\phi_p<0.5$ and $\psi_p>0$ in Figs.~\ref{fig:vector_fields}(a)-\ref{fig:vector_fields}(d).
Regarding the trajectory plotted with a solid (blue) line in Fig.~\ref{fig:trajectory_positive_density}, one part of it from Point $A$ to Point $C$ is not shown due to the difficulty in obtaining an accurate numerical integration near the boundary.
The part of this trajectory from Point $C$ to Point $D$ goes down from Point $C$, makes two turns, and approaches Point $D$. This is also shown by the thinner (blue) arrows at the corresponding places in Figs.~\ref{fig:vector_fields}(b)-\ref{fig:vector_fields}(d).

The fact that Point $D$ is still an attractor in the presence of matter is related to the dynamics of the scale factor $a$. Equation~(\ref{a_dot}) implies that $\dot{a}$ is positive when the Hubble parameter $H$ is positive. Then the matter density keeps decreasing in the evolution and asymptotically comes to zero. Correspondingly, the phase flows approach Point $D$.

The flows with $\rho_m<0$ are between or outside of the two branches of the vacuum constraint surface. They also connect the critical points as the flows with $\rho_m>0$ do. However, the trajectories for $\rho_m<0$ are not shown in this paper because they are not physically meaningful.

Some slices of the vector fields $\{\dot{\phi}_{P},\dot{\pi}_{P}\}$ with $H_P<0$ are shown in Figs.~\ref{fig:vector_fields}(g)-\ref{fig:vector_fields}(i).
The typical behavior of the phase flows can be analyzed in a similar way as in the case of $H_P>0$, and is not included.
%%%%%%%%%%%%%%%%%%%%%%%%%%%%%%%%%%%%%%%%%%%%%%%%%%%%%%%%%%%%%
\section{The cosmological evolution of the $R\ln R$ model \label{sec:evolution_RlnR}}
%%%%%%%%%%%%%%%%%%%%%%%%%%%%%%%%%%%%%%%%%%%%%%%%%%%%%%%%%%%%%
\begin{figure*}
  \begin{tabular}{cc}
    \epsfig{file=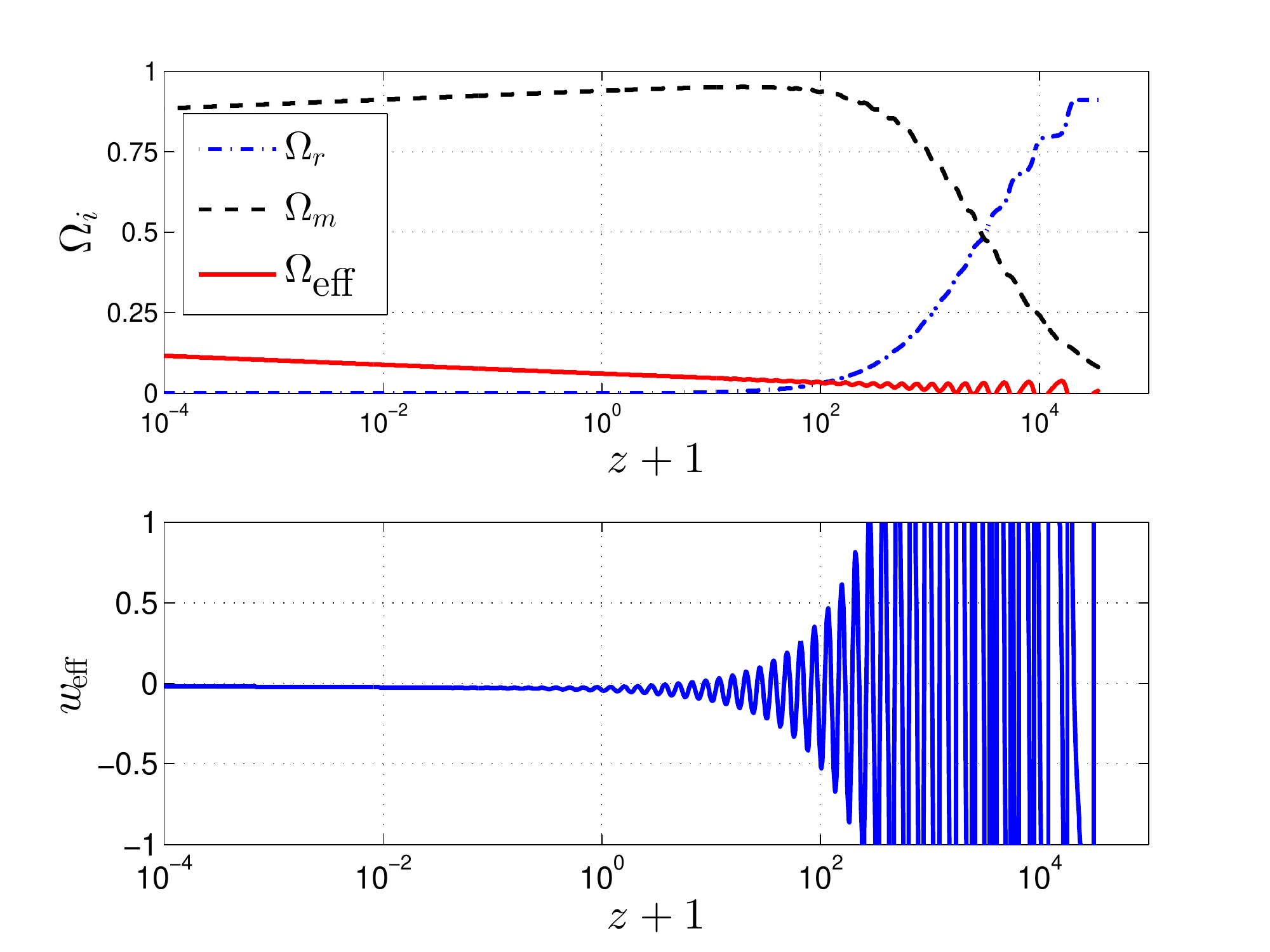, width=9cm} &
    \epsfig{file=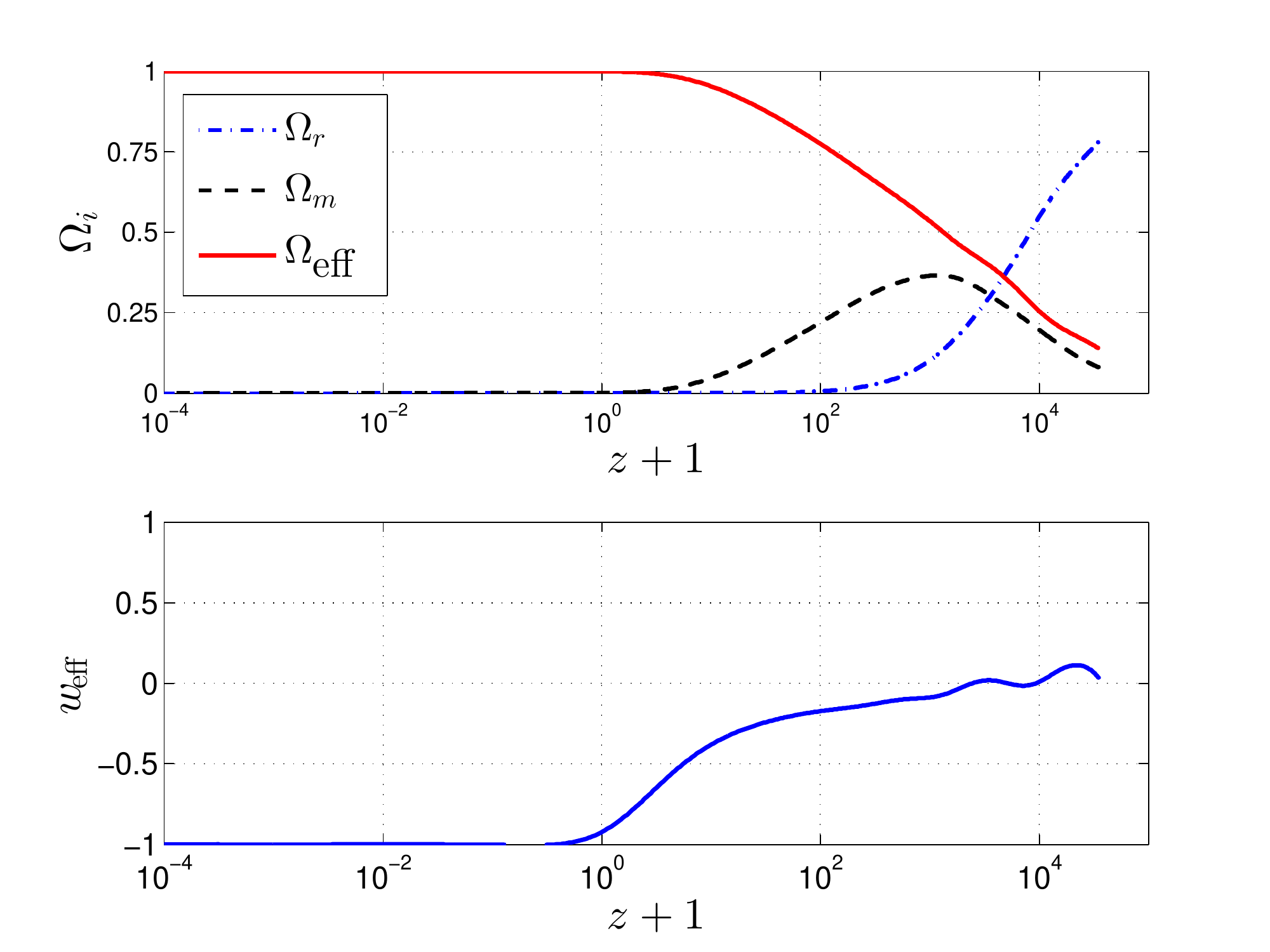, width=9cm} \\
    (a) $\alpha_0=0.002$ &
    (b) $\alpha_0=0.04$ \\
    \epsfig{file=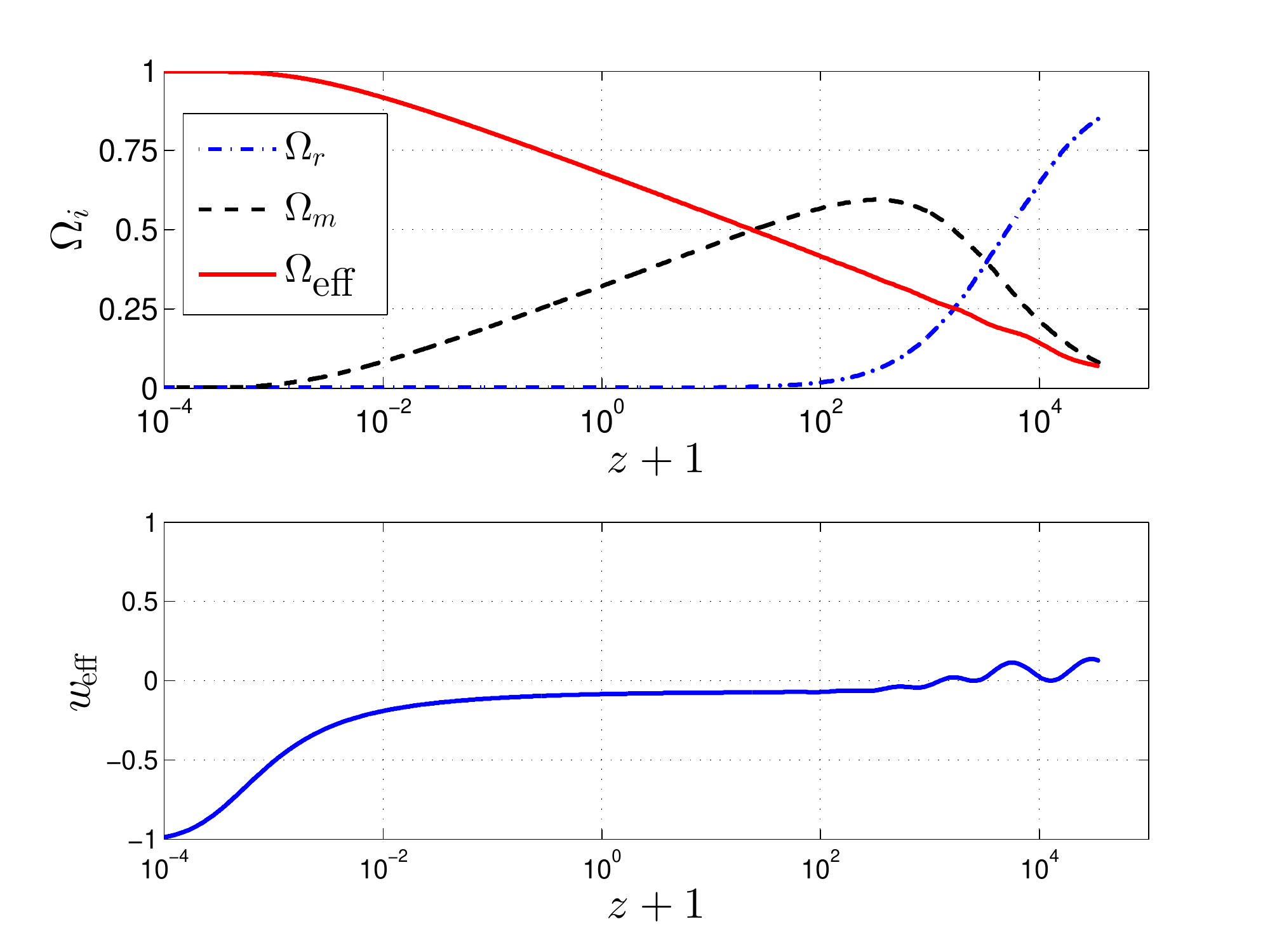, width=9cm} &
    \epsfig{file=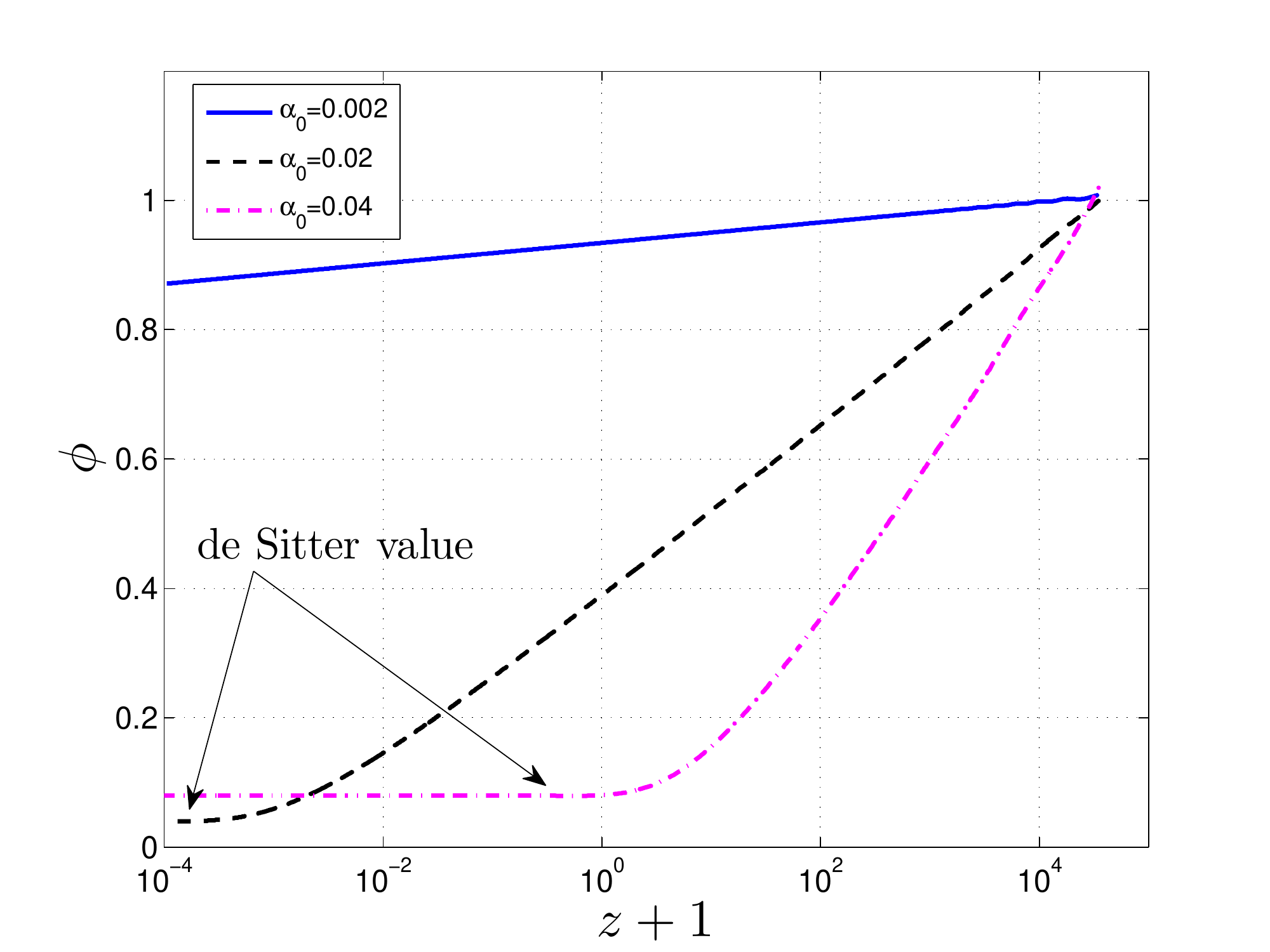, width=9cm} \\
    (c) $\alpha_0=0.02$ &
    (d) evolution of $\phi$\\
  \end{tabular}
  \caption{The cosmological evolution for the $R\ln R$ model with $R_0=1$.
  (a) The cosmological evolution with $\alpha_0=0.002$. A tiny value of $\alpha_0$ will generate an extreme small value of $\Lambda(=R_{0}e^{-1/\alpha_{0}+1})$, then a super long matter-domination stage.
  (b) The cosmological evolution with $\alpha_0=0.04$. A large $\alpha_0$ makes a small $V''(\phi)$ and then a fast evolution of $\phi$.
  (c) The cosmological evolution with $\alpha_0=0.02$.
  (d) The evolution of the field $\phi$ with $\alpha_0$ taking different values.}
  \label{fig:phi_vs_alpha}
\end{figure*}

In the previous section, we studied the global behavior of the phase-space dynamics in $f(R)$ cosmology, where $\rho_m$ and $\rho_r$ are independent of the Hubble parameter, $H$. In this section, we explore the physically more important solution where the scalar field $\phi$ tracks the matter density evolution, and $\rho_m$ and $\rho_r$ are related to $H$ by
\be \frac{8\pi G}{3}(\rho_m+\rho_r)\equiv(\Omega_{m}+\Omega_{r})H^{2}
=\left(\frac{\Omega_{m,0}}{a^{3}}+\frac{\Omega_{r,0}}{a^{4}}\right)H_{0}^{2}, \nonumber\ee
where the \lq\lq$0$\rq\rq~in the indices denotes that the quantities are measured today with $z=0$. The ${\Omega_i}'s$ are defined as $\Omega_{i}=8\pi G\rho_{i}/(3H^2)$, where the index $i$ refers to radiation or matter. At high redshift, the field $\phi$ of this solution closely follows the minimum of the effective potential $V_{\text{eff}}$, which is defined by $V'_{\text{eff}}=V'(\phi)-8\pi G\rho_m/3$ [see Eq.~(\ref{pi_dot})], until the field $\phi$ becomes very light and \lq\lq releases,\rq\rq~approaching the de Sitter minimum of the potential $V(\phi)$.

Equation~(\ref{f_R_logR}) shows that the model is reduced to general relativity when $R$ is equal to $R_0$. However, as argued in Sec.~\ref{sec:introduction_RlnR}, the field $\phi$ logarithmically runs with respect to $X=8\pi G\rho/\Lambda$, and the $R\ln R$ model slowly deviates from general relativity. In this paper, we set units so that $R_0$ is equal to $1$, and let $R$ be equal to $R_0$ around $z=3.5 \times 10^{4}$, where the matter-radiation equality takes place~\cite{Komatsu}.

The cosmological acceleration is a low-curvature issue. Consequently, $f(R)$ gravity should be reduced to general relativity at the high-curvature scale, and it only deviates from general relativity at the low-curvature scale. However, in the $R\ln R$ model, the modification term is not negligible at both the high- and the low-curvature scales. In order to reduce this model to general relativity at the high-curvature scale, the parameter $\alpha_0$ should be much less than 1, as discussed in Sec.~\ref{sec:introduction_RlnR}. However, $\alpha_0$ cannot be too small because of the relation between the de Sitter curvature and $\alpha_0$, $\Lambda=R_{0}e^{-1/\alpha_{0}+1}$, and also because of the relation between the mass of the field and $\alpha_0$. Note that the mass of the field $\phi$ is defined by
\be
m^2\equiv V''(\phi)=\frac{1}{3}\Lambda e^{\phi/\alpha_0-2} \left(\frac{\phi}{\alpha_0}-1\right).
\label{v_dub_prime_RlnR}
\ee
A tiny $\alpha_0$ generates an extremely small value for $\Lambda$ and a heavy mass for the field $\phi$. Consequently, the matter-domination stage would last too long due to the extremely small value of $\Lambda$, and the evolution of $\phi$ would be very slow due to its heavy mass, which is shown in Figs.~\ref{fig:phi_vs_alpha}(a) and \ref{fig:phi_vs_alpha}(d). With the same arguments, the parameter $\alpha_0$ cannot be too large either. A large $\alpha_0$ would result in a short matter-domination epoch (if such an epoch were to exist) and a fast evolution of $\phi$. These are illustrated in Figs.~\ref{fig:phi_vs_alpha}(b) and \ref{fig:phi_vs_alpha}(d), respectively. Consequently, one needs to choose an intermediate value for $\alpha_0$. Letting $\alpha_0$ take the value of $0.02$, we plot the evolution of the ${\Omega_i}'s$ and $w_{\text{eff}}$ in Fig.~\ref{fig:phi_vs_alpha}(c) and that of $\phi$ in Fig.~\ref{fig:phi_vs_alpha}(d), respectively. In this solution, matter-radiation equality takes place around a redshift of $z=3250$~\cite{Komatsu}, and $\Omega_m=0.32$ at $z=0$. The field $\phi$ runs significantly depending on the matter density, as shown in Eq.~(\ref{phi_RlnR}). Consequently, the effective dark energy density also changes significantly as implied in Eq.~(\ref{rho_eff}). As a result, there is no ideal matter-domination epoch at high redshift. Moreover, $w_{\text{eff}}$ is far away from the expected value of $-1$ in the late Universe for this model.

The equation of state $w_{\text{eff}}$ oscillates as shown in Fig.~\ref{fig:phi_vs_alpha}. This is related to the initial conditions in the numerical simulations, and can be explained as follows. At high redshift, the field $\phi$ oscillates near the minimum of the effective potential $V_{\text{eff}(\phi)}$, defined by $V'_{\text{eff}(\phi)}=V'(\phi)-8\pi G\rho_{m}/3$ [see Eq.~(\ref{phi_evolution})]. The close dependence of $w_{\text{eff}}$ on the kinetic terms of $\dot{\phi}$ and $\ddot{\phi}$ also makes $w_{\text{eff}}$ oscillate~\cite{Starobinsky,Tsujikawa4}. [See Eqs.~(\ref{w_eff_tilde_1}), (\ref{rho_eff}), and (\ref{p_eff}), which define $w_{\text{eff}}$.]

In Sec.~\ref{sec:conditions_viability}, we analyzed the cosmological viability of $f(R)$ gravity, and concluded that a heavy mass for the field $\phi$ would result in a slow evolution of $\phi$ and thus a long matter-domination epoch, and vice versa. These conclusions are verified by the evolution of the ${\Omega_i}\rq s$ and $\phi$, shown in Fig.~\ref{fig:phi_vs_alpha}.

\section{Phase-space dynamics of the Hu-Sawicki model\label{sec:dynamics_Hu_Sawicki}}
%%%%%%%%%%%%%%%%%%%%%%%%%%%%%%
\subsection{Introduction to the Hu-Sawicki model}
In the $R\ln R$ model, general relativity is recovered only for a particular range of curvature scales due to the logarithmic running of $f'$ with respect to the matter density. This makes it hard for the $R\ln R$ model to have sensible cosmological evolution. Actually, this problem is alleviated in the modified logarithmic model \cite{Frolov}
\be f(R)=R\frac{a+\ln (R/R_0)}{1+\ln (R/R_0)}=R\left[1-\frac{b}{1+\ln (R/R_0)}\right], \label{modified_log_model}\ee
where $b=1-a$, because in this model $f'$ asymptotes to a finite value and general relativity is restored at the high-curvature scale. In this model, the running of the beta function, $\beta=-k(\alpha-\alpha_{*})^{2}$, is essentially the same as in the $R\ln R$ model. $k$ and $\alpha_{*}$ are positive constants. Therefore, this model can still generate a hierarchy as discussed in Sec.~\ref{sec:introduction_RlnR}. However, the function $f(R)$ still deviates noticeably from general relativity from $R=R_0$ to $R\gg R_0$. In the $\Lambda$CDM-like models, the scalar field $f'$ is almost frozen when the Ricci scalar is higher than the cosmological constant scale, and is released when the Ricci scalar is near the cosmological constant scale. In the rest of this paper, we apply the techniques developed above to a typical example of the $\Lambda$CDM-like models: the Hu-Sawicki model. The function $f(R)$ in this model reads~\cite{Hu_Sawicki}
\be f(R) = R - R_0\frac{C_1R^{n}}{C_2R^{n}+R_{0}^{n}}, \nonumber \ee
where $C_1$ and $C_2$ are dimensionless parameters, $R_0=8\pi G\bar{\rho}_0/3$, and $\bar{\rho}_0$ is the average matter density of the current Universe.
We consider one of the simplest versions of this model, i.e. for $n=1$,
\be f(R) = R-\frac{CR_0 R}{R+R_0}, \label{f_R_Hu_Sawicki}\ee
where $C$ is a dimensionless parameter. With this choice,
\vspace{-8 pt}
\be f'=1-\frac{CR_{0} ^2}{(R+R_0)^2}, \label{f_prime_Hu_Sawicki}\ee
\vspace{-8 pt}
\be R=R_{0} \left[\sqrt{\frac{C}{1-f'}}-1\right], \label{R_Hu_Sawicki}\ee
\vspace{-8 pt}
\be V'(\phi)=\frac{R^{3}}{3(R+R_0)^{2}}\left[1+(1-C)\frac{R_0}{R}\left(2+\frac{R_0}{R}\right) \right]. \label{V_prime_Hu_Sawicki}\ee
Equations (\ref{f_prime_Hu_Sawicki}) and (\ref{V_prime_Hu_Sawicki}) show that as long as the matter density is much greater than $R_0$, the curvature $R$ will trace the matter density well, $\phi$ will be close to $1$ but will not cross $1$, and general relativity will be restored. As implied in Eq.~(\ref{V_prime_Hu_Sawicki}), in order for this model to have a de Sitter attractor where $V'(\phi)=0$, the parameter $C$ needs to be greater than $1$.
In this paper, we set $C=1.2$.

\begin{figure}%[!htbp]
\includegraphics[width=7.5 cm]{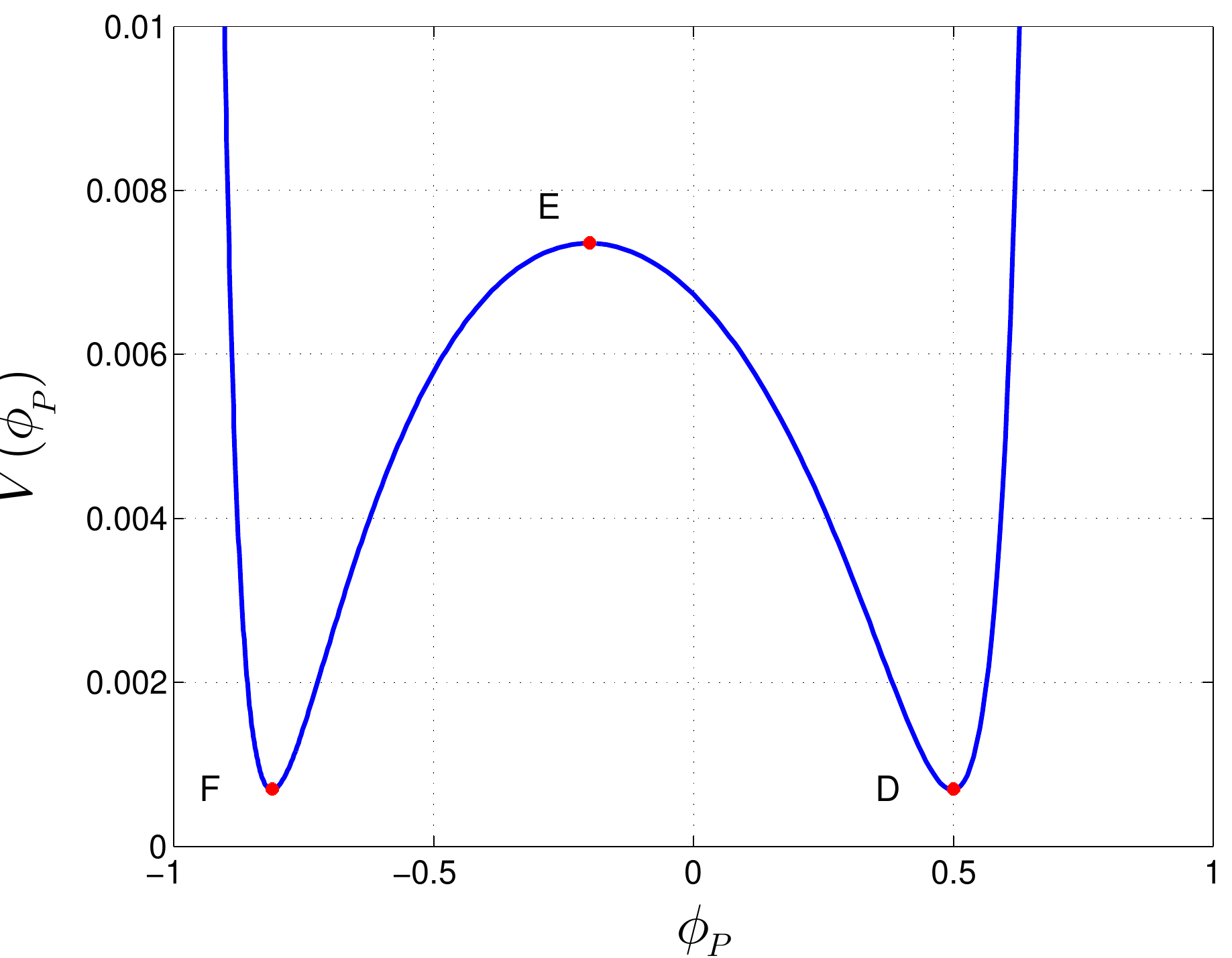}
\caption{The potential $V(\phi_P)$, obtained via integration of Eq.~(\ref{V_prime_Hu_Sawicki}), for the Hu-Sawicki model with $C=1.2$ and $R_0=1$. $\phi_P$ is a compactified coordinate obtained via the $\text{Poincar\'e}$ transformation, $\phi_p=\phi/\sqrt{1+\phi^2}$.}
\label{fig:potential_Hu_Sawicki}
\end{figure}

\begin{figure*}
  \includegraphics[width=17 cm]{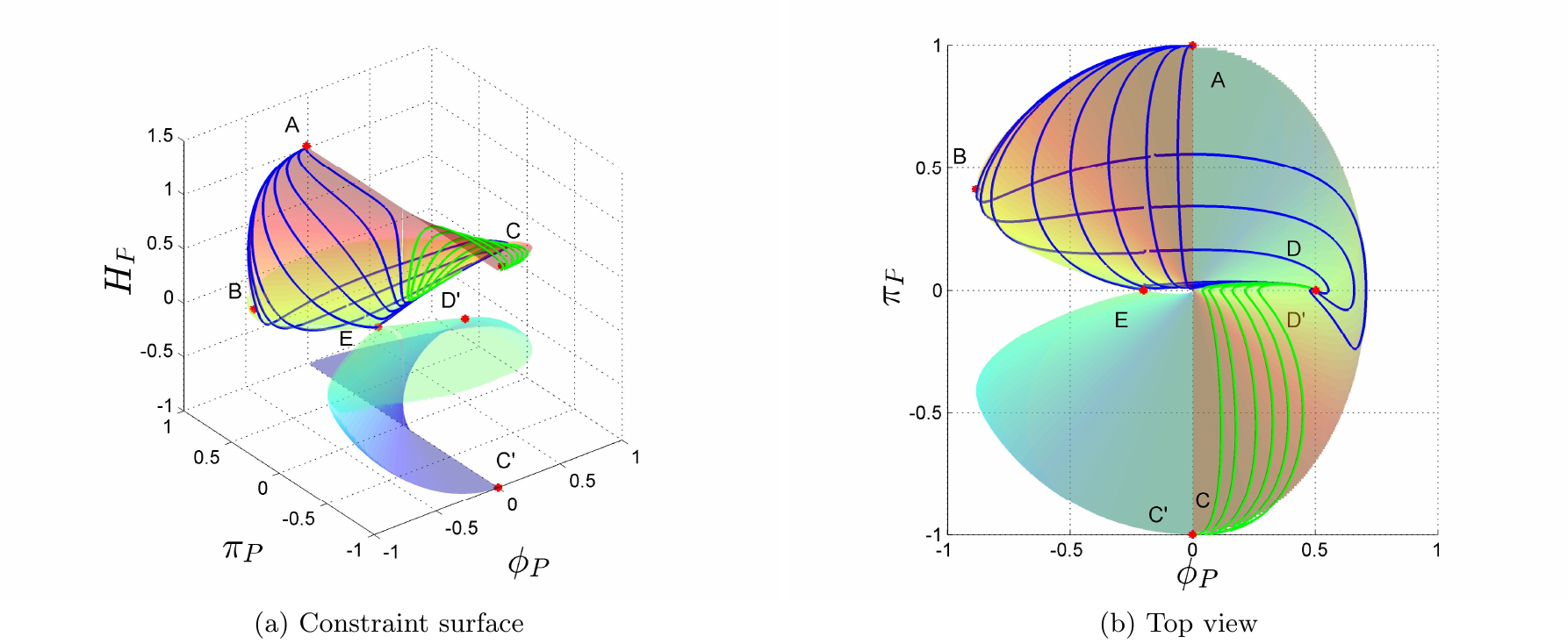}
  \caption{(Color online) The constraint surface and the phase-space flows with $\rho_m=0$ for the Hu-Sawicki model with $C=1.2$ and $R_0=1$. As in the $R\ln R$ model, Point $A$ is a repeller, Points $B$ and $C$ are saddle points, and Point $D$ is an attractor.
  Point $C'$ is an attractor, and Point $D'$ is a repeller.
  Point $E$ is a critical saddle point. It is the lowest point of the $H_P\ge0$ branch of the vacuum constraint surface. It is also the only point connecting the two branches, $H_P\ge0$ and $H_P\le0$, of the constraint surface. In (b), the left boundary of the constraint surface is defined by $\phi\equiv f'=1$. Regarding the trajectories in green (light color), the parts of them between $A$ and $C$ are not plotted due to the difficulty in obtaining accurate numerical integration near the boundary. The shadings correspond to the values of $H_P$. A color bar is not shown because the values of $H_P$ can be seen from the $z$ axis.}
  \label{fig:surface_and_diagram_Hu_Sawicki}
\end{figure*}

Integrating Eq.~(\ref{V_prime_Hu_Sawicki}) leads to the potential $V(\phi)$, as plotted in Fig.~\ref{fig:potential_Hu_Sawicki}, with the integration constant being set arbitrarily. The potential has three critical points: Points $D$, $E$, and $F$. Points $D$ and $F$ are de Sitter points, and Point $E$ is a saddle point. Points $D$ and $E$ are also shown on the constraint surface in vacuum case, as plotted in Fig.~\ref{fig:surface_and_diagram_Hu_Sawicki}. However, Point $F$ is not shown in Fig.~\ref{fig:surface_and_diagram_Hu_Sawicki}, because its corresponding Hubble parameter is a complex number.
\begin{figure*}[t!]
  \hspace{-18pt}
  \begin{tabular}{ccc}
    \epsfig{file=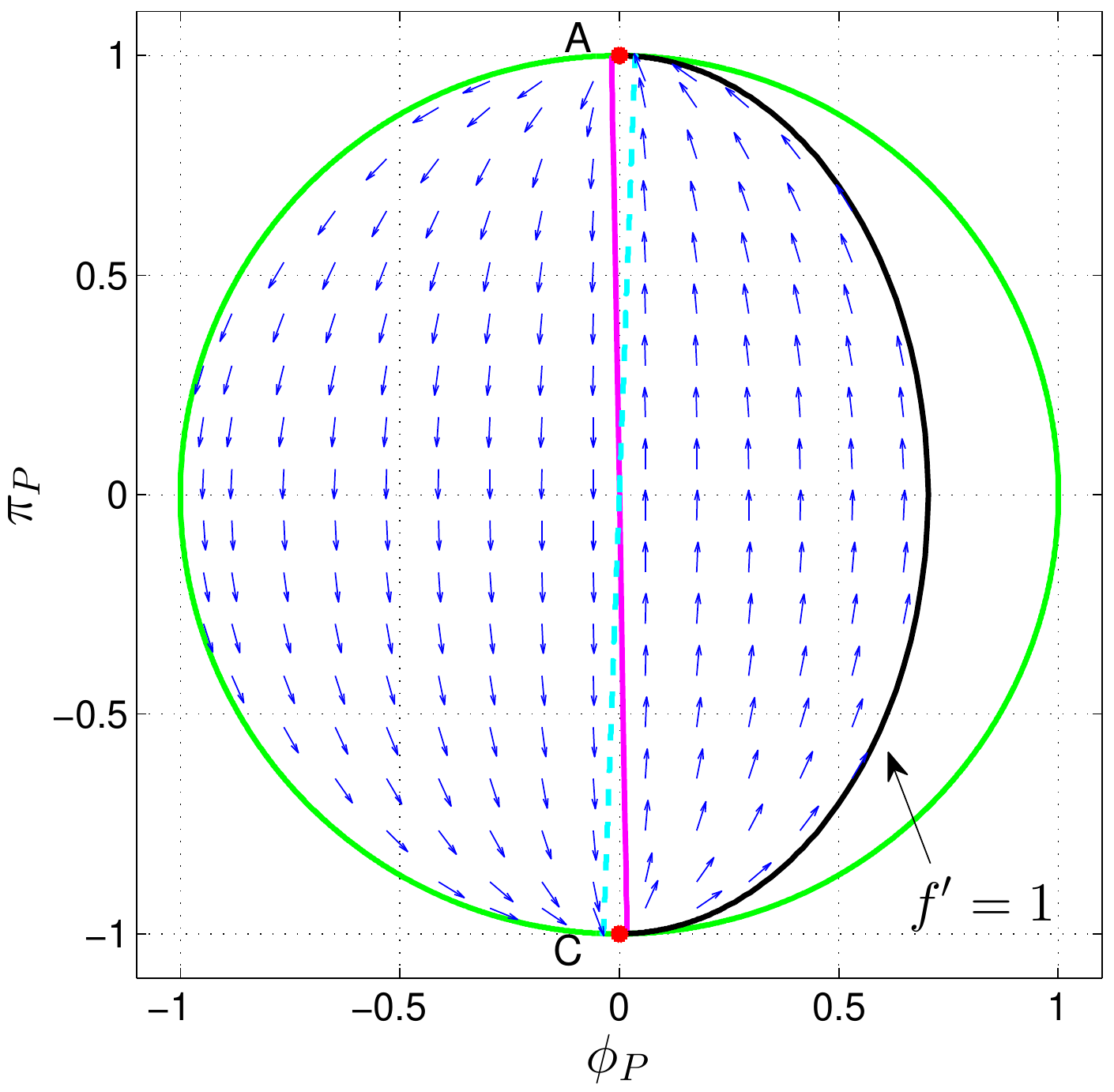, width=5.93cm} &
    \epsfig{file=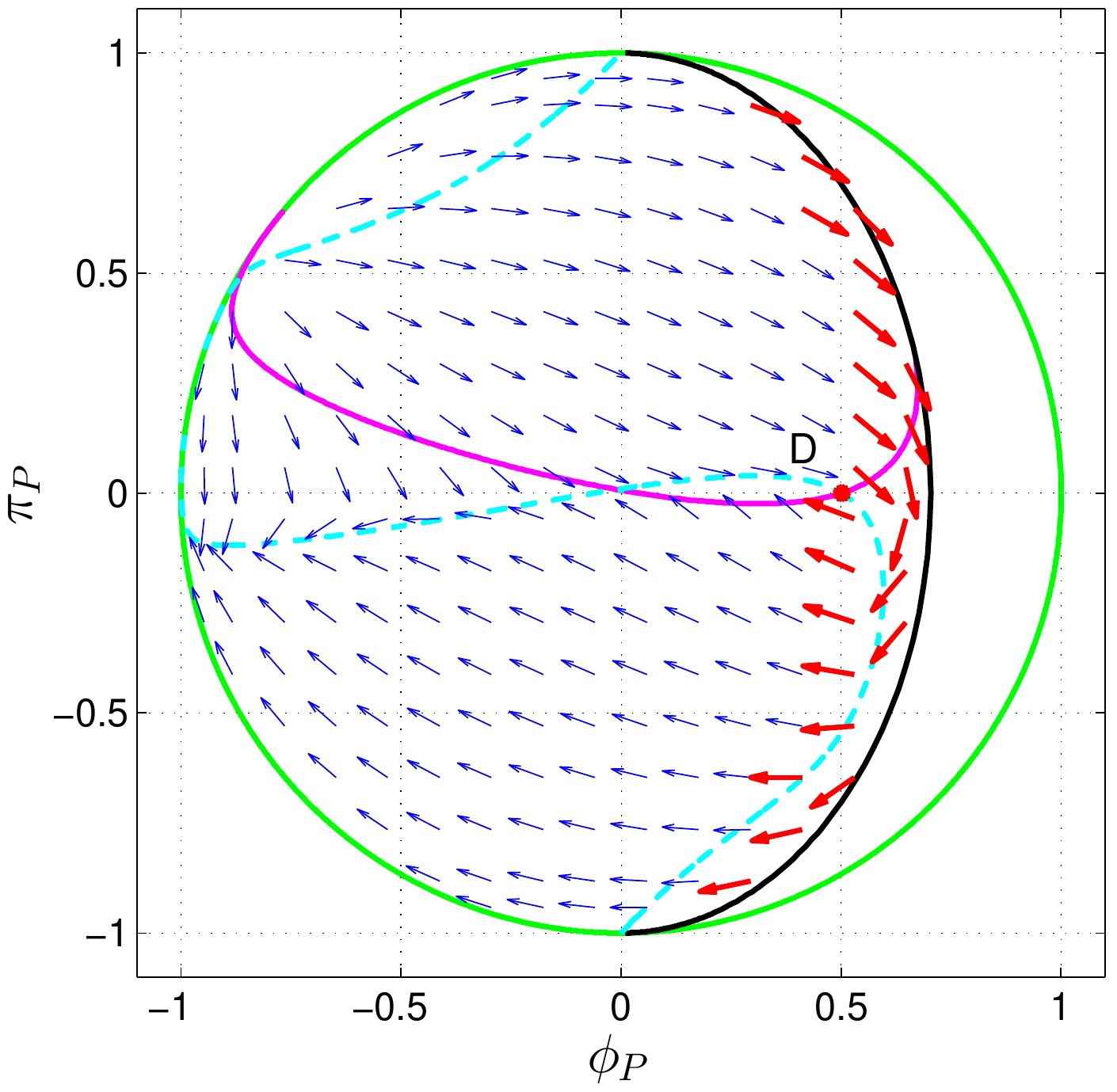, width=5.98cm} &
    \epsfig{file=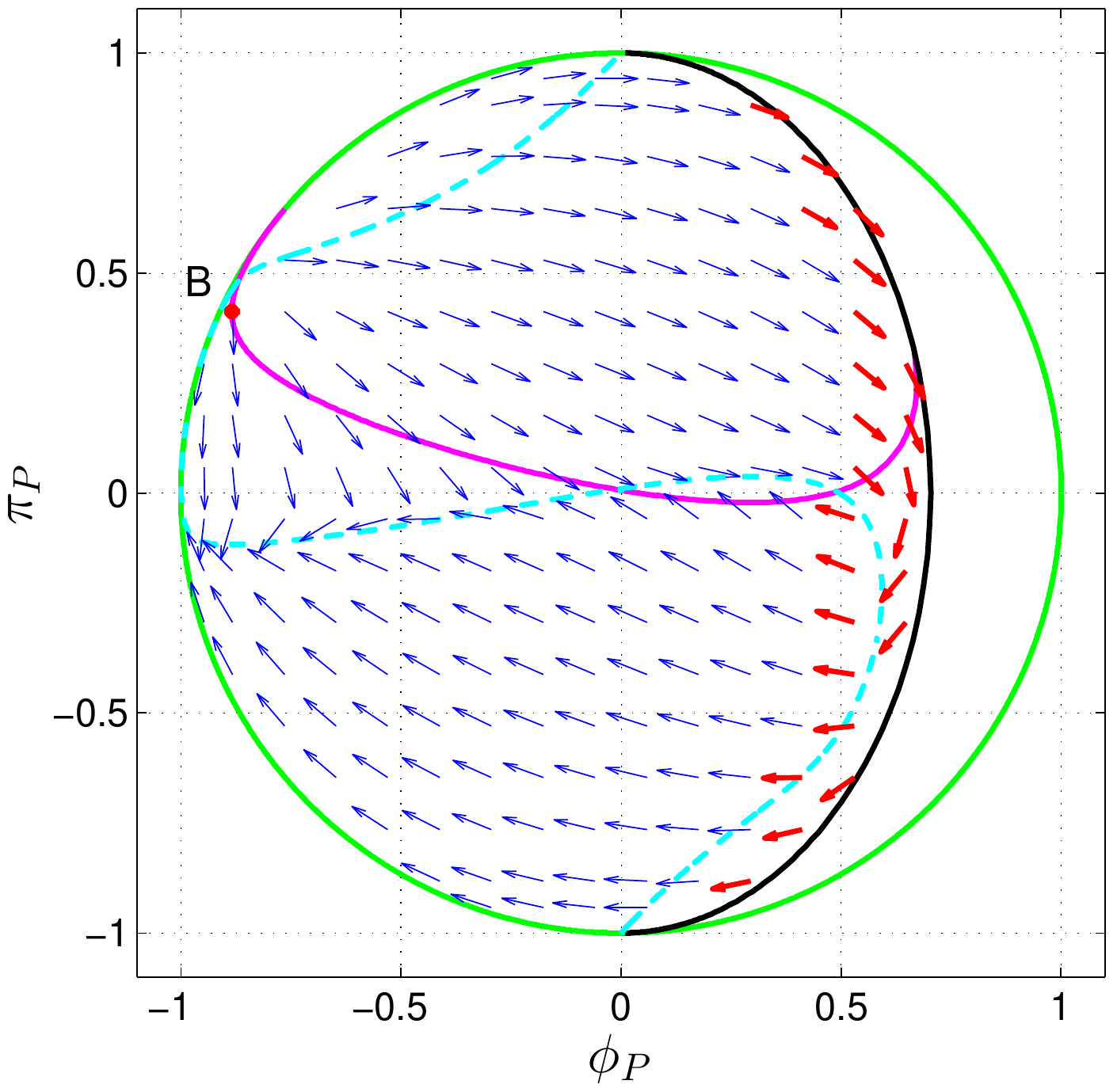, width=6cm} \\
    (a) $H_P\rightarrow1$ &
    (b) $H_P=0.233$ &
    (c) $H_P=0.227$ \\
    \epsfig{file=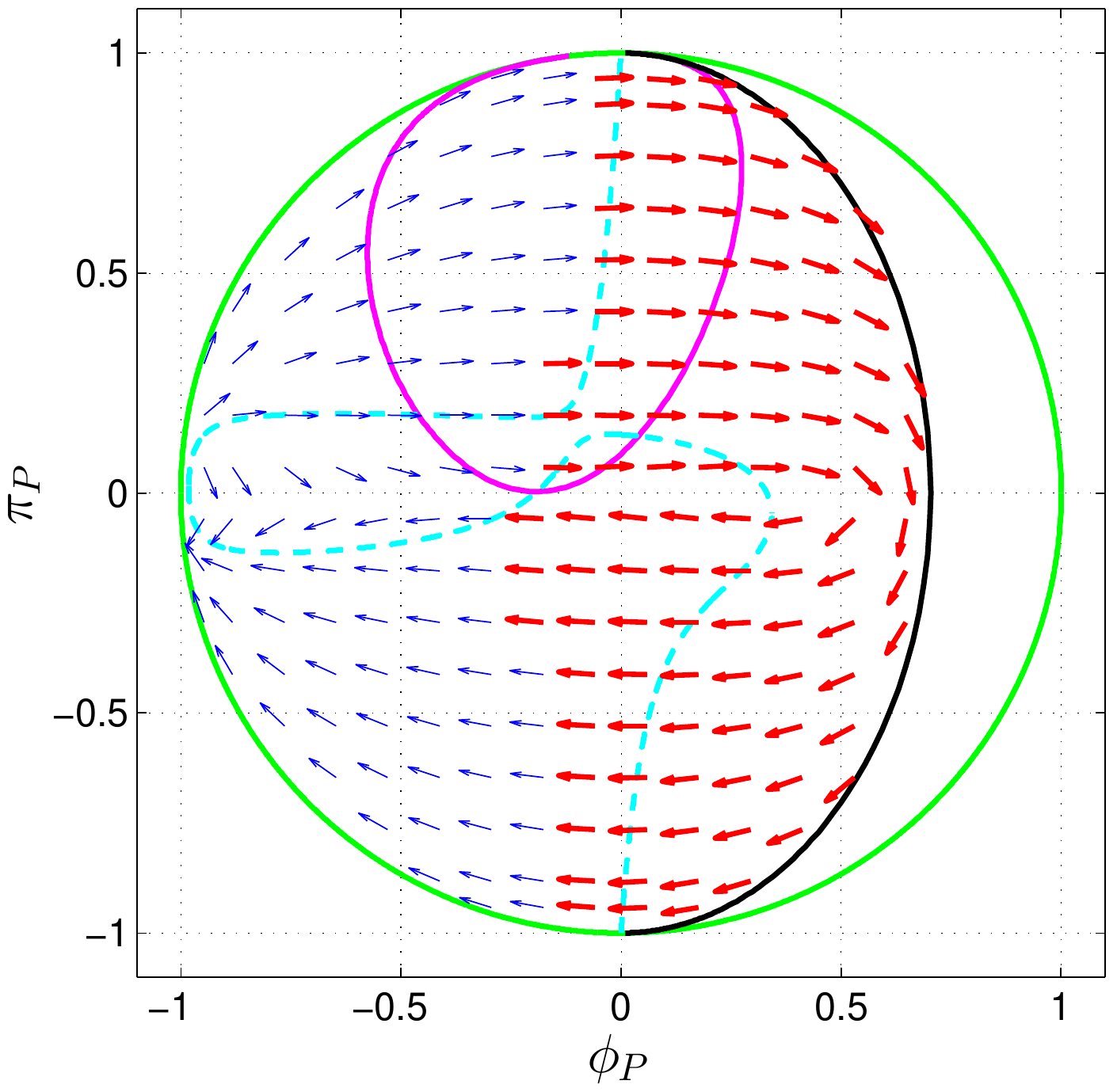, width=6cm} &
    \epsfig{file=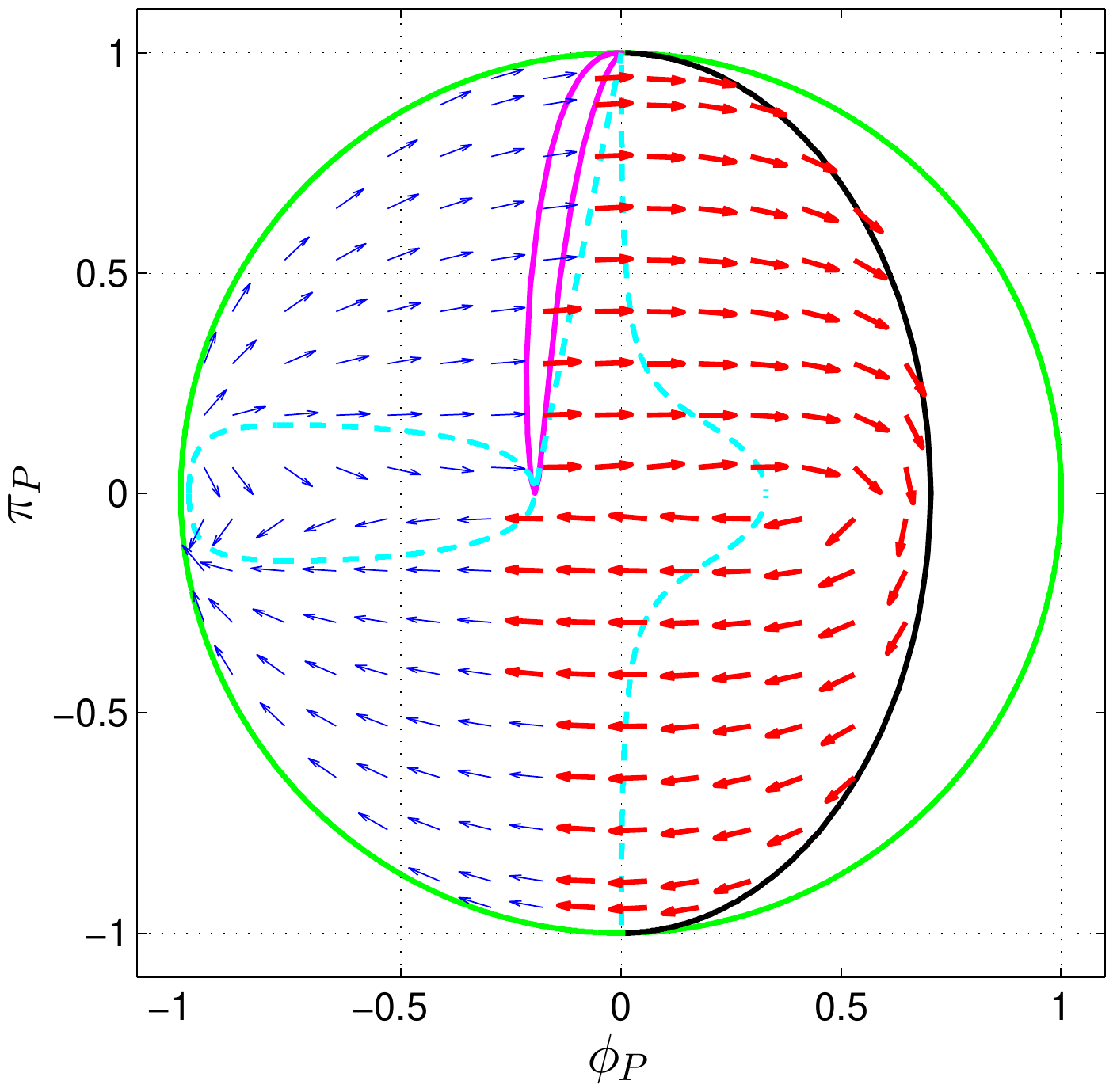, width=6cm} &
    \epsfig{file=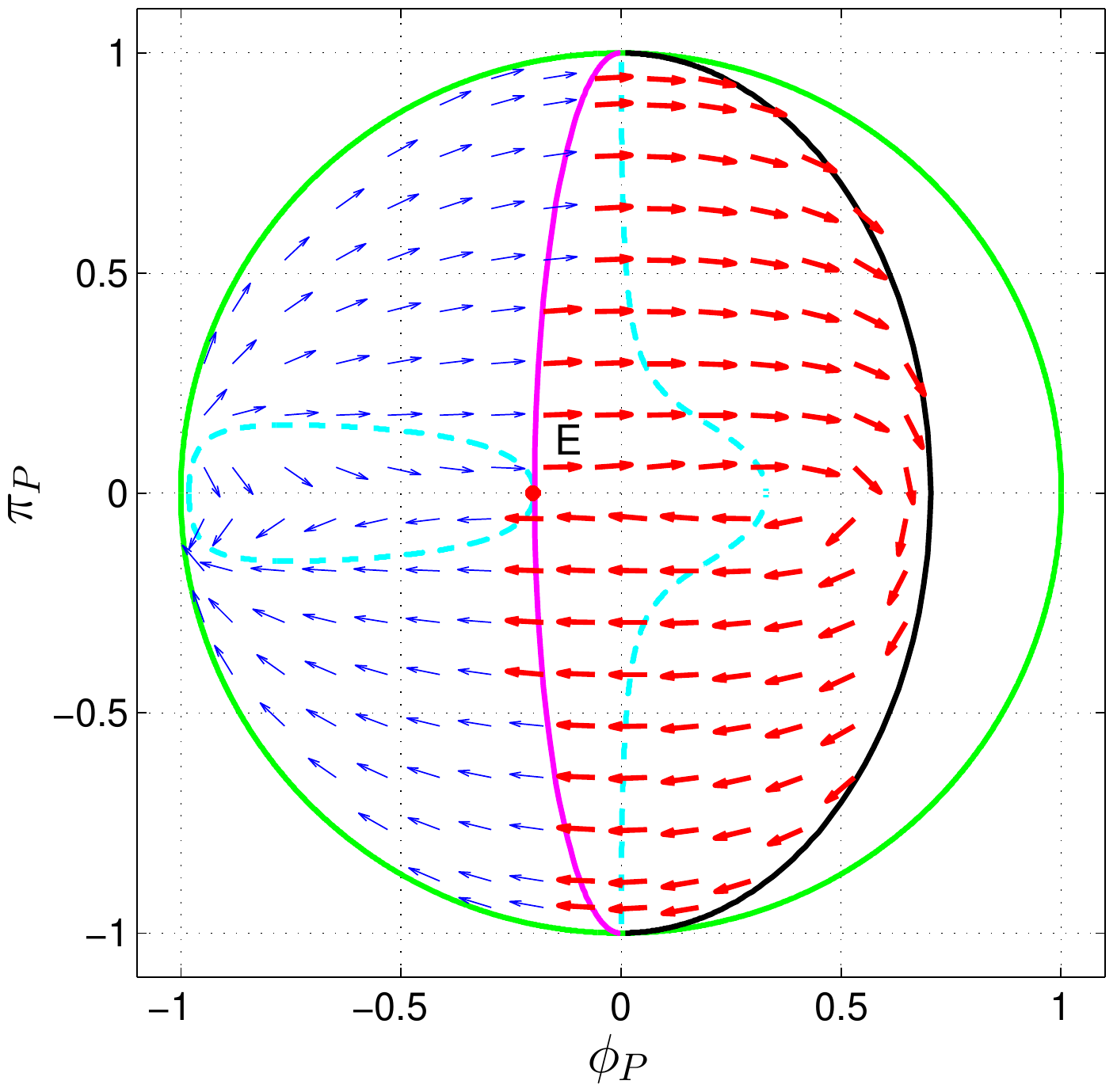, width=6cm} \\
    (d) $H_P=0.017$ &
    (e) $H_P=0.0001$ &
    (f) $H_P=0$ \\
    \epsfig{file=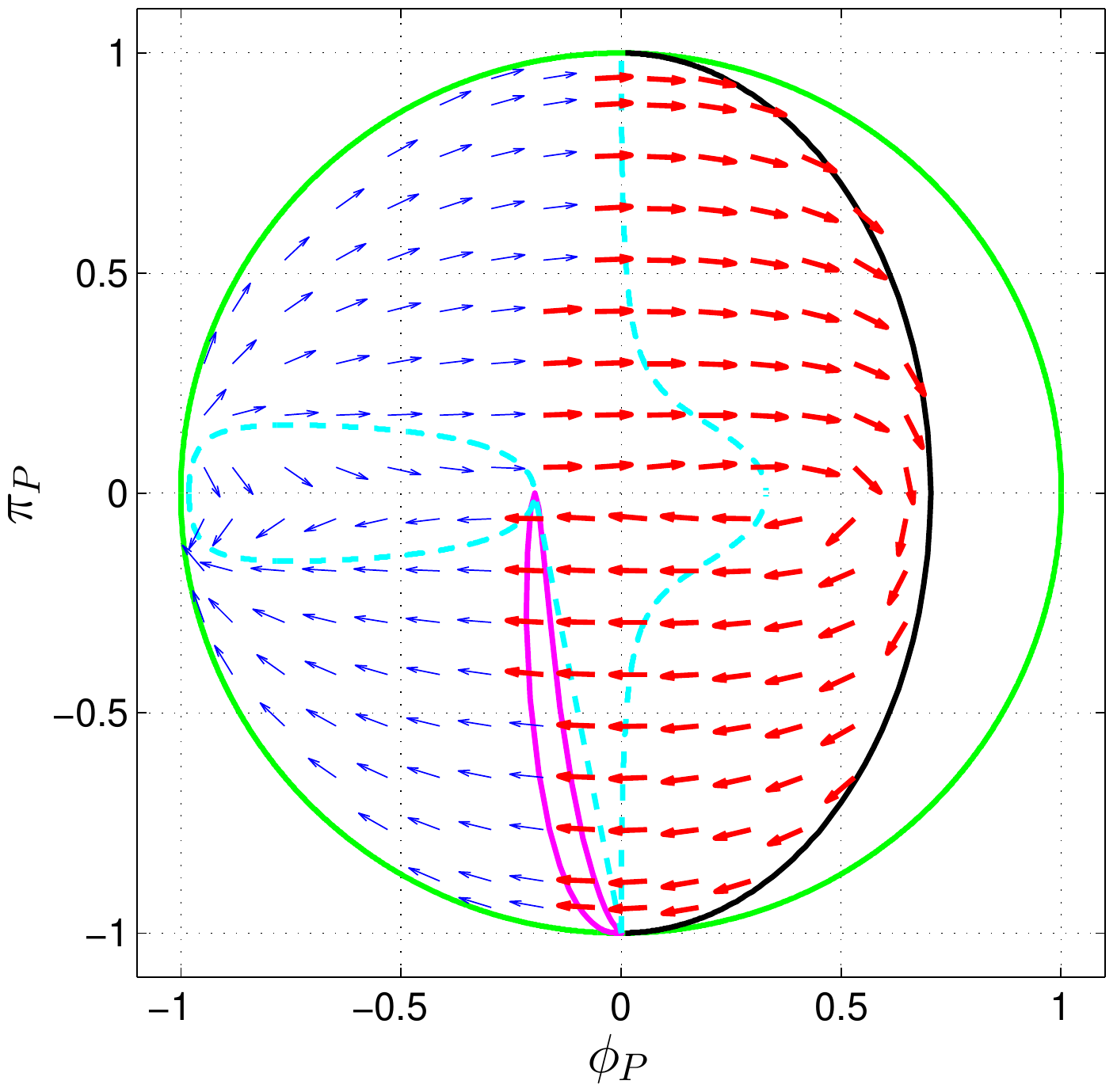, width=6cm} &
    \epsfig{file=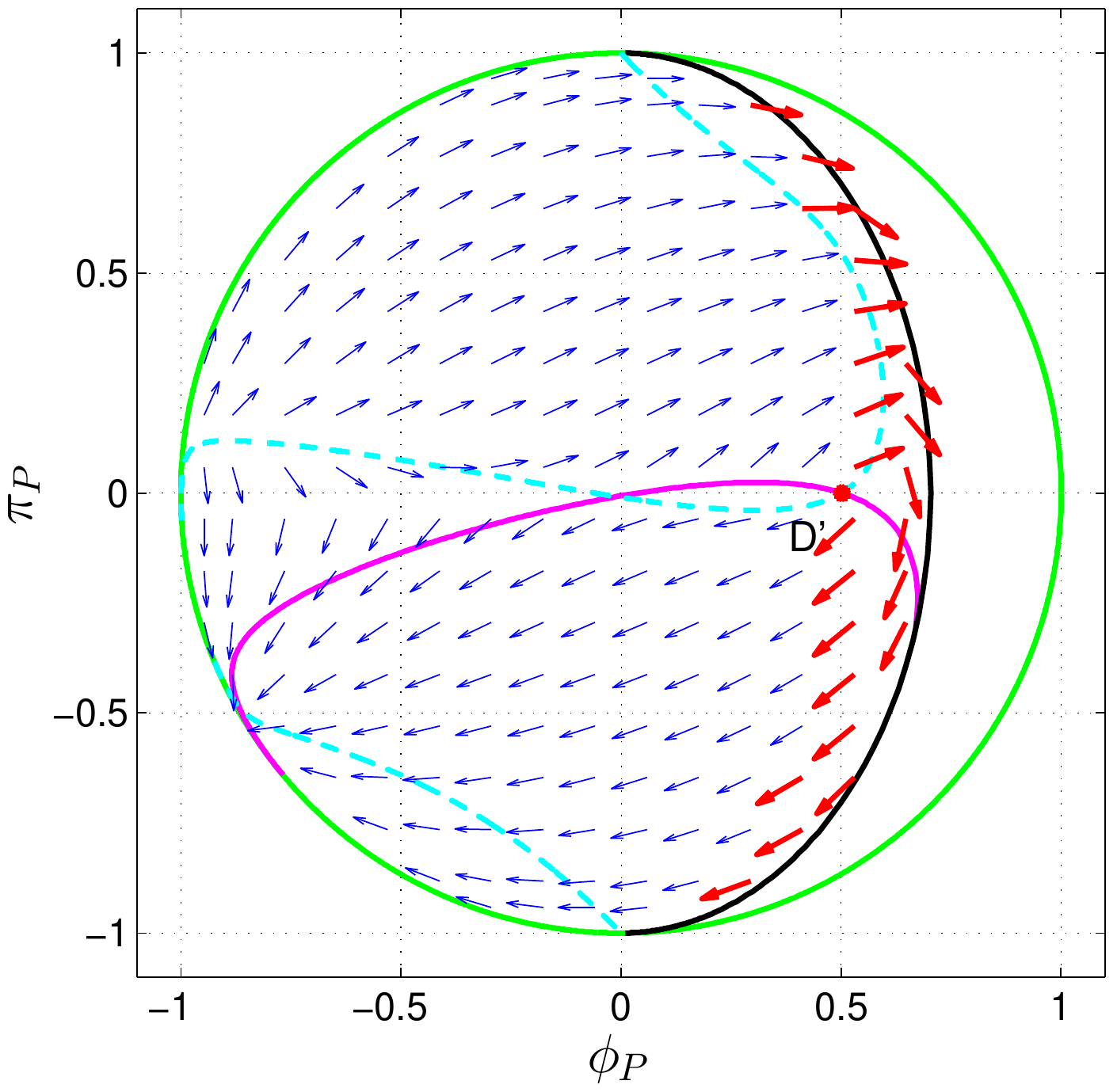, width=6cm} &
    \epsfig{file=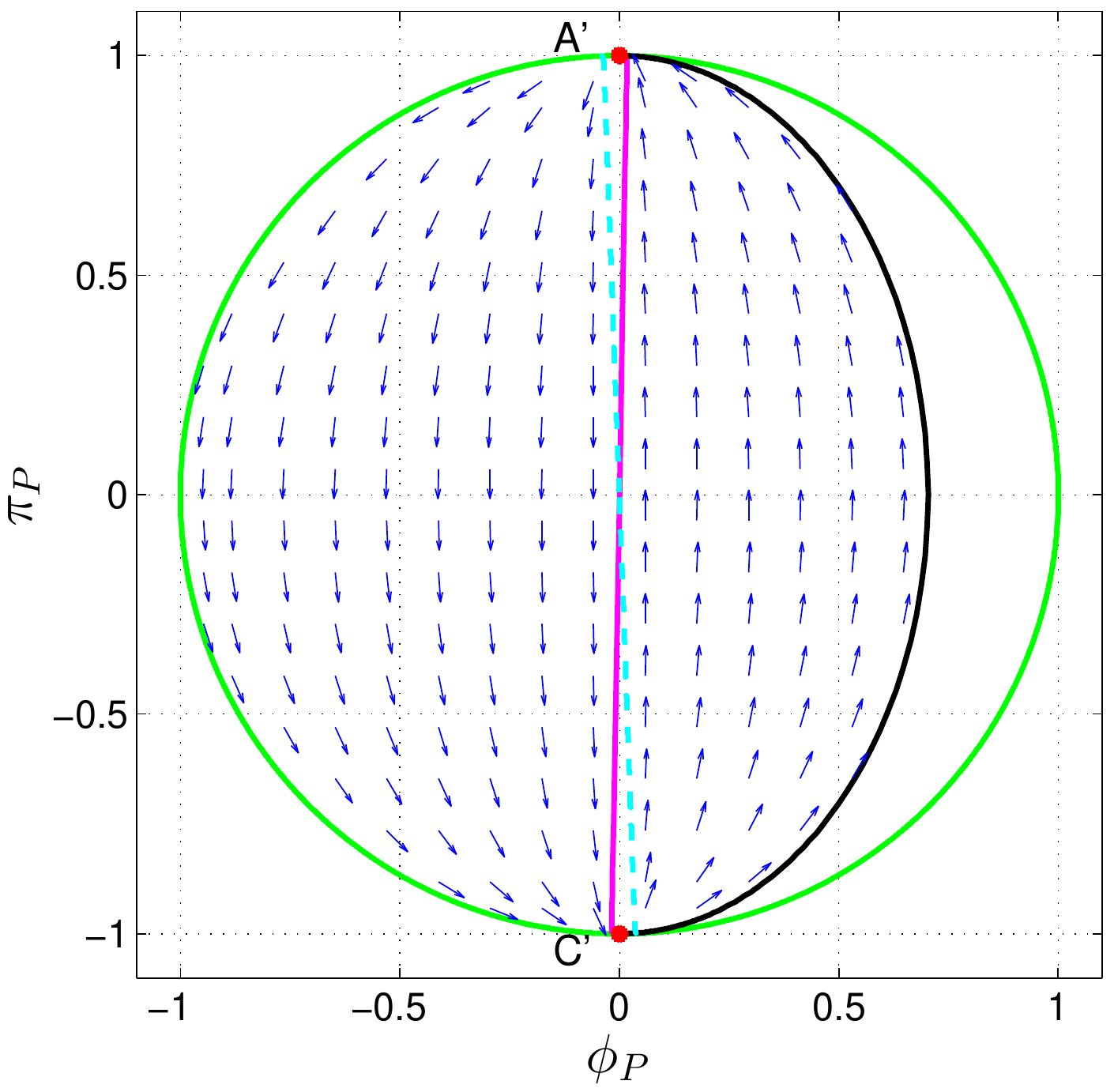, width=6cm} \\
    (g) $H_P=-0.0001$ &
    (h) $H_P=-0.233$ &
    (i) $H_P\rightarrow -1$ \\
  \end{tabular}
  \caption{(Color online) The phase flows on the slices of $H_P=\mbox{const}$ for the Hu-Sawicki model with $C=1.2$ and $R_0=1$. The thinner (blue) arrows denote that $\dot{H_P}<0$. The thicker (red) arrows are for $\dot{H}_{P}>0$. The solid (magenta) line is the intersection between the two-dimensional constraint surface and the slice of $H_P=\text{const}$. The dashed (cyan) line is the trace of $\dot{\pi}_{P}=0$. The solid (black) line is defined by $\phi\equiv f'=1$.
  In (a), Point $A$ is a repeller and Point $C$ is a saddle point.
  In (b), Point $D$ is an attractor.
  In (c), Point $B$ is a saddle point.
  In (f), Point $E$ is a saddle point and the solid (magenta) line is a critical saddle line.
  In (h), Point $D'$ is a repeller.
  In (i), Point $A'$ is a saddle point and Point $C'$ is an attractor.
  }
  \label{fig:vector_fields_Hu_Sawicki}
\end{figure*}

\subsection{Phase-space dynamics in vacuum}
We explore the phase-space dynamics of the Hu-Sawicki model using the $\text{Poincar\'e}$ compactification in Eq.~(\ref{Poincare_transform}). For this model, the parameter $\sigma$ in Eq.~(\ref{Poincare_transform}) is set to $1$, and Eq.~(\ref{f_prime_Hu_Sawicki}) implies that the left boundary of the phase space is constrained by $\phi\equiv f'=1$.

We first study the structure of the vacuum constraint surface, which is plotted in Fig.~\ref{fig:surface_and_diagram_Hu_Sawicki}. It is similar to that in the $R\ln R$ model. The surface is folded in the octants of $(\phi\le0, \pi\ge0, H\ge0)$ and $(\phi\le0, \pi\le0, H\le0)$. There are five critical points on the $H\ge0$ branch of the constraint surface for this model. When $C$ and $R_{0}$ take the values of $1.2$ and $1$, respectively, the coordinates of these critical points are as follows.
\begin{equation}
\begin{aligned}
& A:(\phi_{P}=0^{-}, \pi_{P}=1, H_{P}=1);\\
& B:(\phi_{P}=-0.884, \pi_{P}=0.412, H_{P}=0.227);\\
& C:(\phi_{P}=0^{+}, \pi_{P}=-1, H_{P}=1);\\
& D:\left(\phi_{P}=0.502, \pi_{P}=0, H_{P}=0.233\right);\\
& E:\left(\phi_{P}=-0.196, \pi_{P}=0, H_{P}=0\right).
\end{aligned}
\nonumber
\end{equation}

Similar to the $R\ln R$ model, in the Hu-Sawicki model, Point $A$ is a repeller, Points $B$ and $C$ are saddle points, and Point $D$ is an attractor. Point $E$ is a new critical saddle point. It is on the cutting edge and also on one end of a critical line. The details on Point $E$ and the critical line are discussed below.

In the three-dimensional phase space $\{\phi,\pi,H\}$, the two-dimensional plane $\phi=1-C$ is very special. From Eqs.~(\ref{f_R_Hu_Sawicki}) and (\ref{R_Hu_Sawicki}), one can see that on this plane
\be f(R)=R=0. \label{R_zero_Hu_Sawicki}\ee
The intersections between the constraint equation (\ref{constraint_eq}) in vacuum and the plane $\phi=1-C$ in the octant $(\phi\le0, \pi\ge0, H\ge0)$ can be expressed as follows. On the vacuum $H_{+}$ branch,
\be \phi=1-C, \pi=\pi, H=-\frac{\pi}{\phi},
\label{critical_line_upper}
\ee
and on the vacuum $H_{-}$ branch,
\be \phi=1-C, \pi=\pi, H=0.
\label{critical_line_lower}
\ee
The vacuum $H_{+}$ and $H_{-}$ branches are obtained from the constraint equation (\ref{constraint_eq}) by setting $\rho_m=\rho_r=0$
\be
H_{\pm} = \frac{1}{2}\left[-\frac{\pi}{\phi}\pm \sqrt{\left(
\frac{\pi}{\phi}\right)^{2} - \frac{2(f-\phi R)}{3\phi}} \right].
\nonumber \ee
The line given by Eq.~(\ref{critical_line_lower}) is a critical saddle line. With Eqs.~(\ref{pi_definition})-(\ref{constraint_eq}) and (\ref{R_zero_Hu_Sawicki}), one can see that, on this line, $\dot{\pi}=\dot{H}=0$. Consequently, $\dot{H_P}$ is equal to zero on the transformed line in the compactified phase space $\{\phi_P, \pi_P, H_P\}$. The critical line is the bottom of the $H_P\ge0$ branch of the vacuum constraint surface. The corresponding critical line on the $H_P\le0$ branch is the top of that branch. These two lines are connected by Point $E$, for which $(\phi=1-C, \pi=0, H=0)$. At this point, $\dot{\phi}=\dot{\pi}=\dot{H}=0$, and then $\dot{\phi}_P=\dot{\pi}_P=\dot{H}_P=0$. Point $E$ is a critical saddle point, as shown in Figs.~\ref{fig:potential_Hu_Sawicki} and~\ref{fig:surface_and_diagram_Hu_Sawicki}. Moreover, Point $E$ is also the only point connecting the two branches of $H_P\ge0$ and $H_P\le0$ of the vacuum constraint surface, as shown in Fig.~\ref{fig:surface_and_diagram_Hu_Sawicki}. The combined critical line is shown by a solid (magenta) line in Fig.~\ref{fig:vector_fields_Hu_Sawicki}(f).

The dynamics on the constraint surface of $H_P<0$ can be explored in a similar way, and is omitted.

\subsection{Phase-space dynamics in the presence of matter}
Some typical slices for the vector fields of $\{\dot{\phi}_{P},\dot{\pi}_{P}\}$ with $H_P$ taking different values in the presence of matter are shown in Fig.~\ref{fig:vector_fields_Hu_Sawicki}. In the case of $\rho_m>0$, the phase flows in the Hu-Sawicki model move similarly to those in the $R\ln R$ model. Specifically, the flows start from Point $A$, and end at Point $D$. The phase flows for $\rho_m<0$ are skipped because they are not physical.

The phase-space dynamics of the modified logarithmic model given by Eq.~(\ref{modified_log_model}), the exponential model given by Eq.~(\ref{exponential_model}) (below)~\cite{Cognola,Linder,Bamba,Elizalde}, and the Tsujikawa (hyperbolic tangent) model given by Eq.~(\ref{Tsujikawa_model}) (below)~\cite{Tsujikawa4} are also analyzed here:
\be f(R)=R-b[c-\exp(-R/R_0)], \label{exponential_model}\ee
\vspace{-18pt}
\be f(R)=R-b\tanh(R/R_0). \label{Tsujikawa_model}\ee
In Eqs.~(\ref{exponential_model}) and (\ref{Tsujikawa_model}), $b$, $c$, and $R_0$ are constants. When the models have a de Sitter attractor with the parameters taking appropriate values, the phase-space dynamics is similar to that of the Hu-Sawicki model. The noticeable differences are given below. In the modified logarithmic model,
\be f'=1-\frac{b}{1+\log (R/R_0)}+\frac{b}{[1+\log (R/R_0)]^{2}}\ge 1-\frac{b}{4}. \nonumber\ee
Therefore, the left boundary of $\phi\equiv f'$ in the phase space of $\{\phi,\pi,H\}$ is constrained by $\phi\ge 1-b/4$. In the Tsujikawa model described by Eq.~(\ref{Tsujikawa_model}), the left boundary of $\phi$ in the phase space of $\{\phi,\pi,H\}$ is constrained by
\be \phi\equiv f'=1-(b/R_0)\sech ^{2}(R/R_0)\ge 1-b/R_0. \nonumber\ee
A critical point, labeled as $E$ and at which $f(R)=R=0$, is present in the Hu-Sawicki model. However, a similar point is absent in the modified logarithmic model described by Eq.~(\ref{modified_log_model}) and the exponential model described by Eq.~(\ref{exponential_model}). This occurs because in the modified logarithmic model the Ricci scalar $R$ cannot be zero due to the logarithmic function shown in $f(R)$ and $f'$ of this model, and in the exponential model $f(R)|_{R=0}=-b\cdot c \ne 0$.

\subsection{Cosmological evolution}

In Sec.~\ref{sec:conditions_viability}, the conditions for the existence of a matter-domination epoch in the early Universe for $f(R)$ gravity were explored, which can be expressed as the requirement that the corrections should be less than the main terms at three orders of derivative with respect to the Ricci scalar $R$. Namely, if we rewrite the function $f(R)$ as $f(R)=R+A(R)$, with $A(R)$ being the modification term, then we have
\be |A(R)|\ll R, \mbox{ } |A'(R)|\ll 1, \mbox{ and } A''(R)\ll 1/R. \label{condition_modification_term}\ee
We can now compare the cosmological evolution of the $f(R)$ models discussed above with the results of Eq.~(\ref{condition_modification_term}).

For the $R\ln R$ model expressed by Eq.~(\ref{f_R_logR}), we have
\be A''(R)=\frac{\alpha_0}{R}.\nonumber\ee
As discussed in Sec.~\ref{sec:evolution_RlnR}, under the balance of general relativity restoration in the early Universe and dark energy domination in the late Universe, a value around $0.02$ is chosen for $\alpha_0$. In this case, $A'(R)$ does not run slowly enough with respect to $R$ to ensure an ideal matter-domination epoch.
For the modified logarithmic model (\ref{modified_log_model}), we have
\be A''(R)=\frac{b}{R}\left\{\frac{1}{[1+\ln(R/R_{0})]^{2}}-\frac{2}{[1+\ln(R/R_{0})]^{3}}\right\}.\nonumber\ee
Therefore, in the early Universe where $R$ is much greater than $R_{0}$, $A''(R)$ in this model runs more slowly than the one in the $R\ln R$ model, and hence has a better cosmological evolution, as shown in Ref.~\cite{Guo}.
The $\Lambda$CDM-like models (e.g., the Hu-Sawicki model) are very close to the $\Lambda$CDM model at high redshift. At low redshift, the modification term in the function $f(R)$ becomes important, and the dark energy is dominant enough to drive the cosmic acceleration. In one of the simplest versions
of the Hu-Sawicki model (\ref{f_R_Hu_Sawicki}),
\be A''(R)=\frac{2CR_0}{(R+R_0)^{3}}\ll \frac{1}{R}, \mbox{ when $R\gg R_0$.}\nonumber \ee
Thus, $A'(R)$ moves more slowly with respect to $R$ than the one in the modified logarithmic model. Consequently, models of this type better fit the cosmological observations in both the early and the late Universe.
The exponential model (\ref{exponential_model}) and the Tsujikawa (hyperbolic tangent) model (\ref{Tsujikawa_model}) are similar in terms of $A''(R)$ to the Hu-Sawicki model, and they have similar cosmological evolution as well~\cite{Bamba}.

%%%%%%%%%%%%%%%%%%%%%%%%%%%%%%%%%%%%%%%%%%%%%%%%%%%%%%%%%%%%%%%%%%%%%%%%%%%%%%%%%%%%%%%%%%%%%%%%%%%%%%%%%%%%%%%
\section{Conclusions \label{sec:conclusions}}
%%%%%%%%%%%%%%%%%%%%%%%%%%%%%%%%%%%%%%%%%%%%%%%%%%%%%%%%%%%%%%%%%%%%%%%%%%%%%%%%%%%%%%%%%%%%%%%%%%%%%%%%%%%%%%%
In this article, we studied the cosmological evolution in $f(R)$ gravity, and obtained the conditions of cosmological viability by using the scalar field description of $f(R)$ gravity. In the early Universe, the field $\phi$ is coupled to the matter density, acquiring mass from this coupling; thus it has a slow-roll evolution. Consequently, general relativity is recovered and a matter-domination stage is ensured in the early Universe. In the late Universe, when the scalar curvature is around the cosmological constant scale, the field $\phi$ will be released from its coupling to the matter density and approach the de Sitter minimum of the potential $V(\phi)$. Then, the dark energy will be dominant and drive the cosmic speed-up. The fact that in the early Universe the field $\phi$ slow-rolls is due to the heavy mass obtained from the coupling between the field $\phi$ and the matter density. This behavior is very close to that of the chameleon mechanism explored in the context of the Solar System tests of $f(R)$ gravity.

The phase-space dynamics and the cosmological evolution of the $R\ln R$ model and the Hu-Sawicki model were studied in detail. The $R\ln R$ model has the feature of being singularity-free, which is an advantage in terms of the hierarchy problem between the cosmological acceleration scale and the Planck scale. On the other hand, in this model general relativity can only be restored at a certain high-curvature regime due to the logarithmic running of $f'$ with respect to the matter density. Therefore, it is hard for this model to have a sensible cosmological evolution in the early Universe. The Hu-Sawicki model is very close to the $\Lambda$CDM model, and can generate a cosmological evolution compatible with the observations of both the early and the late Universe.

In our explorations of phase-space dynamics, for simplicity the radiation density was set to zero. In order to obtain a global picture of the phase space, we compactified the infinite phase space into a finite space via the $\text{Poincar\'e}$ transformation. The $R\ln R$ model and the Hu-Sawicki model have similar phase-space dynamics.
In the vacuum case where the matter density is zero, the phase space is three-dimensional and the constraint surface is two-dimensional. The phase-space dynamics was explored in the three-dimensional phase space $\{\phi,\pi,H\}$ without difficulty. In the expansion branch of the phase space, the constraint surface has a repeller and a de Sitter attractor; while in the contraction branch, the constraint surface has an attractor and a de Sitter repeller. The phase flows simply move from the repeller to the corresponding attractor in each space.
When the matter density is not zero, the phase space $\{\phi,\pi,H, a\}$ is four-dimensional, and the constraint surface is three-dimensional. For ease of visualization, we projected the four-dimensional phase space $\{\phi,\pi,H, a\}$ onto the three-dimensional phase space $\{\phi,\pi,H\}$ by taking the scale factor $a$ as an implicit variable. It is not convenient to study the phase-space dynamics on the three-dimensional constraint surface directly. Instead, we cut the three-dimensional surface into two-dimensional slices of $H_P=\text{const}$, and explored the vector fields of the phase flows on the slices.

As a supplement, we plotted some typical trajectories of the phase flows. Like those in the vacuum case, when the matter density is not zero, the phase flows still move from the repeller to the corresponding attractor in each space. Some trajectories between the repeller and the attractor are similar to those in the vacuum case, while some others are not.
We also explored the phase-space dynamics of some other $f(R)$ models, such as the modified logarithmic model, the exponential model, and the Tsujikawa model. The results are similar to those of the Hu-Sawicki model. We presented some generic features of the phase-space dynamics in $f(R)$ gravity in this paper.
We developed new techniques to explore the phase-space dynamics: compactifying the infinite phase space into a finite space via the $\text{Poincar\'e}$ transformation; studying the vector fields on the two-dimensional slices of the constraint surface when the constraint surface is three-dimensional; and plotting typical trajectories of the phase flows. These techniques are very general and could be applied to studies of other similar dynamical systems.\\

%%%%%%%%%%%%%%%%%%%%%%%%%%%%%%%%%%%%%%%%%%%%%%%%%%%%%%%%%%%%%%%%%%%%%%%%%%%%%%
\section*{Acknowledgments}\small
This  work  was  supported  by  the  Discovery  Grants  program  of  the  Natural  Sciences  and Engineering Research Council of Canada. The authors would  like  to thank Aaron Plahn, Levon Pogosian, Miguel Quartin, Howard Trottier, and Shaojie Yin  for useful discussions. The authors also thank the referee for the many very helpful comments. J.Q.G. thanks Memorial University of Newfoundland for its hospitality during CCGRRA14.
\appendix*
%%%%%%%%%%%%%%%%%%%%%%%%%%%%%%%
\section{Lambert $W$ function}
%%%%%%%%%%%%%%%%%%%%%%%%%%%%%%%
%
The Lambert $W$ function is defined \cite{Corless} by
\be Y=W(Y)e^{W(Y)}, \label{lambertW_definition} \ee
where $Y$ can be a negative or a complex number. In this paper, we only consider the case of $Y>0$. When $0<Y\ll1$, $W(Y)\ll1$, $e^{W(Y)}\rightarrow 1$, then $W(Y)\approx Y$.
When $Y\gg1$, $W(Y)\gg1$, then $\ln Y \approx W$.
Concisely,
\be
W(Y) = \left\{
\begin{array}{l l}
  Y
  & \quad \mbox{if $0<Y\ll1,$ }\\
  %\\
    \ln Y & \quad \mbox{if $Y\gg1.$ }\\
\end{array} \right.
\label{w_function}
\ee

%%%%%%%%%%%%%%%%%%%%%%%%%%%%%%%%%%%%%%%%%%%%%%%%%%%%%%%%%%%%%%%%
%\bibliographystyle{abbrv}
%\bibliography{main}

%%%%%%%%%%%%%%%%%%%%%%%%%%%%%%%%%%%%%%%%%%%%%%%%%%%%%%%%%%%%%%%%%%%%%%%%%%%%%%%%%%%%%%%%%%%%%%%%%%%%%%%%%%%%%

\end{document}